\documentstyle[epsf]{article}
\begin{document}

\parskip 2mm plus 1mm \parindent=0pt
\def\cl{\centerline}\def\ll{\leftline}\def\rl{\rightline}
 \def\hs{\hskip1mm} \def\h10{\hskip10mm} \def\hx{\h10\hbox}
\def\vs{\vskip3mm} \def\page{\vfill\eject}
\def\<{\langle} \def\>{\rangle} \def\br{\bf\rm} 
 \def\de{\partial} \def\Tr{{\rm Tr}} \def\dag{^\dagger} 
\def\half{{\scriptstyle{1\over 2}}} \def\Im{{\rm Im}} \def\Re{{\rm
Re}} \def\I{{\rm I}}  \def\R{{\rm R}} \def\M{{\rm M}} \def\d{{\rm
d}}  \def\P{{\bf P}} \def\cc{{\rm cc}} \def\HC{{\rm HC}} \def\D{{\br D}}
\def\A{{\rm A}} \def\B{{\rm B}}\def\X{{\rm X}} \def\Y{{\rm Y}}\def\L{{\bf L}}
\def\N{{\cal N}} \def \S{{\rm S}}
 \def\rv{\vec r} \def\a{{\bf a}}
 \def\dxi{\Delta\xi}\def\dpsi{\Delta\psi} \def\dt{\Delta t} \def\ds{\Delta s} 
 \def\dw{\Delta w} \def\dx{\Delta x} \def\du{\Delta u}
\def\ne{=\hskip-3.3mm /\hskip3.3mm}
\def\Pr{{\br Pr}\hskip.3mm}\def\tr{{\rm tr}}\def\Pa{P_{\rm a}}
\def\sgn{{\rm sgn}} \def\to{\rightarrow} \def\Ra{\Rightarrow}
\def\ra{\rightarrow}\def\o{\circ}
\def\Fs{F^{\rm s}} \def\Fsa{F^{{\rm s}1}} \def\Fsb{F^{{\rm s}2}}
\def\I{Id}
\def\CL{\longrightarrow^{\hskip-5mm \rm CL}\hs}
\def\NI{\longrightarrow^{\hskip-5mm \rm NI}\hs}

\rm
icp - 990527
\vskip9mm

\vs \bf
\cl{Quantum measurement breaks Lorentz symmetry}
{\vskip4mm}
\centerline {by}
\vskip4mm \centerline {Ian C Percival}\vskip6mm 
\cl{Department of Physics} \vskip3mm
\centerline {Queen Mary and Westfield College,
University of London}
\vs
Tel: 0044 171 775 3289\h10 Fax: 0044 181 981 9465
\vs
email: i.c.percival@qmw.ac.uk

\vskip4mm {Abstract}\rm

Traditionally causes come before effects, but according to modern
physics things aren't that simple.  Special relativity shows that
`before' and `after' are relative, and quantum measurement is even
more subtle.  Since the nonlocality of Bell's theorem, it has been
known that quantum measurement has an uneasy relation with special
relativity, described by Shimony as `peaceful coexistence'.  Hardy's
theorem says that quantum measurement requires a preferred Lorentz
frame.  The original proofs of the theorem depended on there being no
backward causality, even at the quantum level.  In
quant-ph/9803044 this condition was removed.  It was only required
that systems with classical inputs and outputs had no causal loops.
Here the conditions are weakened further: there should be no {\it
forbidden} causal loops as defined in the text.  The theory depends on
a transfer function analysis, which is introduced in detail before
application to specific systems.

\vfill
\vs
\vs
keywords: quantum, Bell, collapse, measurement, Lorentz, transfer, causality
\rm
\vs

99 May 27, QMW TH 99-07 \hfill To be submitted
\eject

\page

{\obeylines
{\bf 1 Introduction} p3

\vs\vs{\bf 2 Deterministic systems} p4
\hs{2.1 Physical systems with inputs and outputs}
\hs{2.2 Transition and transfer pictures}
\hs{2.3 Combining systems}
\hs{2.4 Causal loops}

\vs\vs{\bf 3 Stochastic systems} p15
\hs{3.1 Transition and transfer pictures}
\hs{3.2 Background variables}
\hs{3.3 Independence}
\hs{3.4 Linked systems and causal loops}

\vs\vs{\bf 4 Space and time} p23
\hs{4.1 Time-ordering and causality}
\hs{4.2 Systems and ports in space and time}
\hs{4.3 Signalling transfer functions}

\vs\vs{\bf 5 Inequalities of the Bell type} p25
\hs{5.1 Quantum experiments in spacetime}
\hs{5.2 Background variables}
\hs{5.3 Bell-type experiments}
\hs{5.4 Bell inequalities}
\hs{5.5 Simplified and moving Bell experiments}

\vs\vs{\bf 6 Hardy's theorem and the double Bell experiment} p31
\hs{6.1 Hardy's theorem}
\hs{6.2 The double Bell experiment}
\hs{6.3 Breaking Lorentz symmetry}

}

\page

\vs{\bf 1 Introduction}

Bell's inequalities are a consequence of classical constraints on
causal relations in spacetime, where here {\it classical} means
non-quantum, as against nonrelativistic.  Constraints such as the Bell
inequality have significance for purely classical stochastic systems,
as well as for quantum measurement theory.  Here they are put in the
context of a general theory of such constraints.  This general theory
follows from an application of a spacetime approach to the
input-output theory of the engineers.  The approach requires only
that space and time should be put on a common basis, and that the
spacetime constraints on causality should be clearly stated in the
context of input-output systems.  The rest, including Bell's theorem,
and the breakdown of Lorentz symmetry, follows.

Bell's ideas on quantum measurement and related problems are put in a
wider context within a general formalism.  In this more general
context, a double Bell thought experiment described here and in
\cite{Percival1998b,Percival1998d} is used to demonstrate Hardy's
theorem that quantum measurements require a preferred Lorentz frame.
The tension between Bell nonlocality and special relativity is well-known
and has been discussed widely, for example by Shimony \cite{Shimony1984}
Helliwell and Konkowski \cite{Helliwell1983} and at length in the book
by Maudlin \cite{Maudlin1994}.

In the famous controversy between Einstein and Bohr on quantum theory,
Einstein found it difficult or impossible to accept nonlocal
interactions.  Bell's theorem shows that nonlocality is unavoidable.
Hardy's theorem goes further by showing that quantum measurement
requires a preferred Lorentz frame.  It does not satisfy the
conditions of Lorentz invariance.  The original proofs of the theorem
depended on there being no backward causality, even at the quantum
level.  The double Bell experiment does not.  It combines two Bell
experiments in a configuration resembling Einstein's original special
relativistic thought experiment with light signals.  When combined
with Lorentz invariance and Bell's theorem, this leads to a forbidden
causal loop with probability greater than zero.  This is ruled out by
a stochastic causal loop constraint, leading either to Hardy's result
or to other consequences that are very difficult to accept.

Causal loops are introduced first for deterministic systems and then
for stochastic systems.  There is a causal loop constraint for
each.  The transfer function theories of the Bell experiment and
the double Bell experiment are presented. 

\page

 \cl{\bf 2 Deterministic systems}

\vs{\bf 2.1 Physical systems with inputs and outputs}

Physical systems may be isolated or they may interact with one
another.  This work develops the theory of those systems whose
interaction with other systems is restricted to discrete classical
inputs and outputs with well-defined properties.  {\it System} here
refers to such a system. 

Such systems are the elements of digital circuits as used in
communications and computer technology.  Many of the ideas presented
here are taken from these fields, but there is an important
difference.  In technology it is conventional and convenient to treat
space and time very differently.  In physics, particularly
relativistic physics, it is better to treat them far as possible on
the same basis.  In engineering, a system is restricted to a localized
region of space, but is not usually considered to be restricted with
respect to time.  An engineering system which {\it is} restricted with
respect to both space and time, for example a communication system
with a signal confined to given intervals of time at the input and
output, is an example of the kind of physical system treated here.
Our physical systems are restricted to regions of spacetime in the
same way that engineering systems are restricted in space.  Of course,
in practice, no engineering system lasts for ever, so all of them are
like our systems, but this finite lifetime is often ignored in the
engineering context.

The spacetime properties of our physical systems are so important that
they are treated separately from the other properties, in order that
the role of the spacetime properties can be seen as clearly as
possible.  So we start by treating systems and their interactions {\it
without} considering where they are in spacetime, and introduce the
spacetime relations and their properties afterwards.

A system may be entirely classical, or partly classical and partly
quantum, but here the inputs and outputs are {\it always} classical,
even though the rest of the system may be quantal.  We treat quantum
measurements and also other types of interaction between quantum systems
and classical systems.  We do not consider purely quantum systems
except as a link between classical inputs and classical outputs.  It
is often helpful to treat these classical and classical-quantum
systems as black boxes, whose only important property is the relation
between the classical input and the classical output.

Every system $S$ contains classical input and output ports, which
contain the inputs and outputs.  We start by describing the properties
of systems with only one input and one output port, whose inputs and
outputs are discrete, with a finite number of possible states.  The
inputs are labelled $i$, with $\N(i)$ possible values, and the outputs
are labelled $j$, with $\N(j)$ possible values.
 
Because the inputs and outputs are classical and discrete, their
values can be measured, that is, they can be monitored, without
affecting the evolution of the system.  For continuous variables this
is not true.  Monitoring continuous variables always produces a
perturbation.  In classical theory the perturbation can be made
arbitrarily small, but there are practical limits and also a quantum
limit beyond which it is not possible to go.  Quantum systems cannot
be monitored without producing major changes in the system, except for
special cases.  These are the reasons for restricting our attention to
systems with discrete classical inputs and outputs.

The simplest nondegenerate ports are binary ports with only two
possible values.  These are usually taken to be 0 and 1, although in
some contexts we use $+$ and $-$ or even 1 and 2.  Systems with only
binary ports are binary systems, and these are the elements from which
digital computers are built.  Much of the theory will be illustrated
using systems with binary ports.

The distinction between input and output ports is made by the
different roles they play in the interaction of a system with its
environment.  Since each system is confined in spacetime, the
environment is the {\it spacetime} environment.  `Another system' in
the spacetime context may occupy the same region of space at a
different time, so from the engineer's viewpoint it is the same
system.

As usual, input and output ports are distinguished by their causal
relations with the system and its environment, which often includes
other such systems.  An input $i$ at the input port of a system $S$ is
determined by its environment outside $S$.  It affects the evolution
of $S$ in general, and the outputs $j$ at the output ports in
particular.  The outputs affect the spacetime environment, but do not
affect the rest of the system.  The system is the physical link by
which the input affects the output, which is its most significant role
here.  This is particularly important when the link between the inputs
and outputs is a quantum system, as in a controlled quantum
measurement.

We are concerned with the evolution of the system as it affects the
causal relations between the inputs and outputs, which may be
deterministic or stochastic.

\vs{\bf 2.2 Transition and transfer pictures}

If $i$ determines $j$ uniquely, then the system is deterministic.
Sections 2.2-2.4 deal with deterministic systems.  Many important
spacetime relations between inputs and outputs, for example `an input
can have no influence on an earlier output' are not treated in detail
here.  They have separate sections 4.1-4.3 to themselves.

A system may have many input and output ports which are separated from
one another.  We then use $i_p$ to represent the input at port $p$ and
$j_q$ the output at port $q$.  The value of all the inputs together is
represented by $i=(i_1,i_2,\dots)$ and $j=(j_1,j_2,\dots)$ represents
the value of all the outputs.  We refer to $i$ and $j$ as the {\it
the} input and {\it the} output.  Both may be distributed among many
ports, but this does not affect our definitions of the causal
relations between the overall input $i$ and the overall output $j$,
which are the same as for elementary systems with only one port of
each kind, such as the one illustrated in figure 2.1a.
\begin{figure}[htb]

\epsfxsize14.0cm
\centerline{\epsfbox{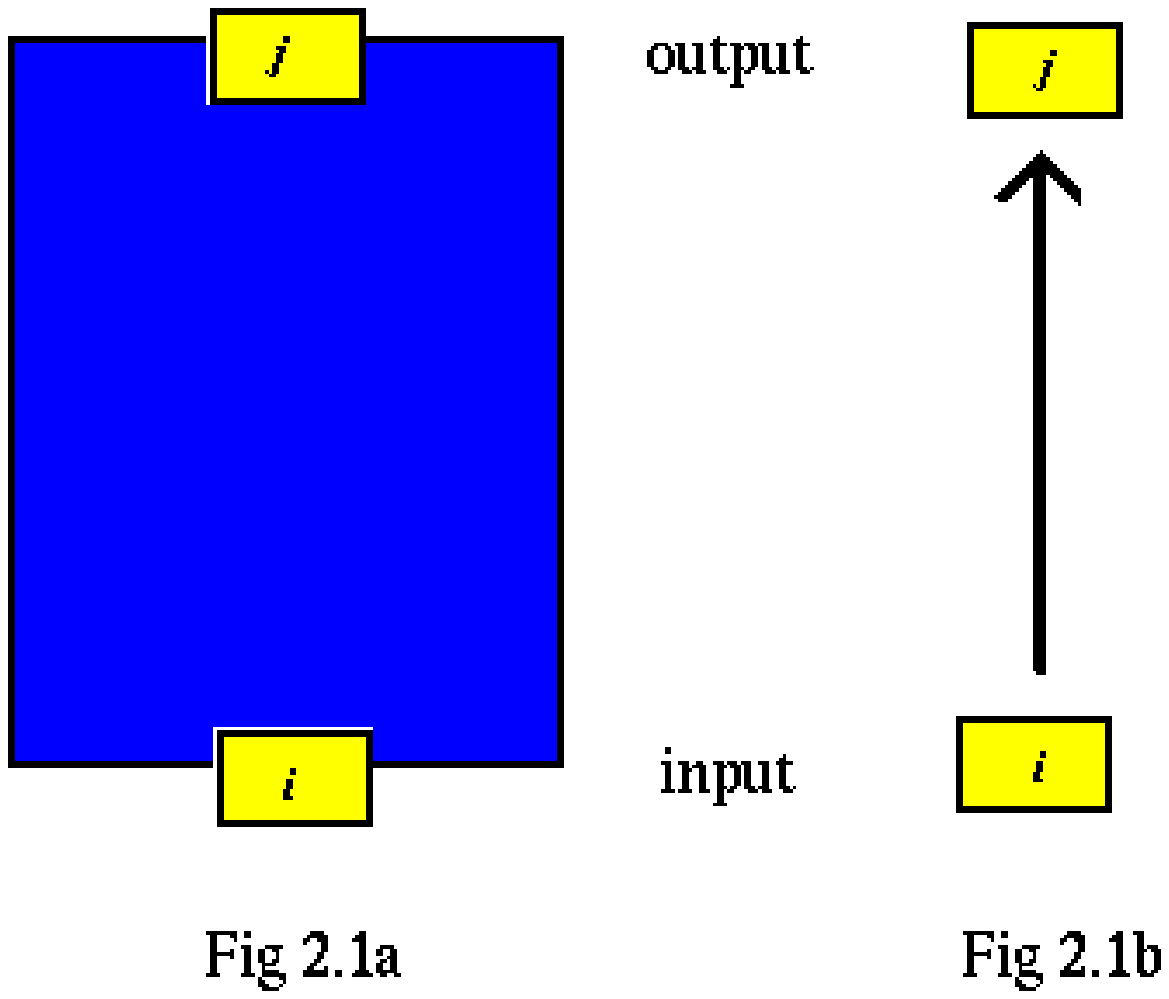} }
\label{fig2.1}

\caption{An elementary system with two ports.}
\end{figure}
Figure 2.1b is an alternative illustration of the system, where the
arrow represents a {\it causal link}.  This means that $j$ depends on
$i$.  If $j$ is independent of $i$, then the arrow is absent,
and the same applies to the inputs and outputs at particular
ports $i_p$ and $j_q$.

The number of possible values of $i$ and of $j$ are $\N(i)$ and
$\N(j)$.  For every value of the input $i$ there is a corresponding
value of the output $j$, and the relation between them is the {\it
transition} $i\to j$.  The transitions for all inputs $i$ together can
expressed as a function, the {\it transfer function}
$$
F\hx{such that}\h10 j=F(i).
\eqno(2.1)$$
We shall often treat the system as a black box, in which case the
dependence of the outputs on the inputs, given by the transfer
function, defines all the properties of a deterministic system that
concern us.  Either the transitions $i\to j$ for every $i$, or the
single transfer function $F$ give the black box properties of the
system.
\begin{figure}[htb]

\epsfxsize14.0cm
\centerline{\epsfbox{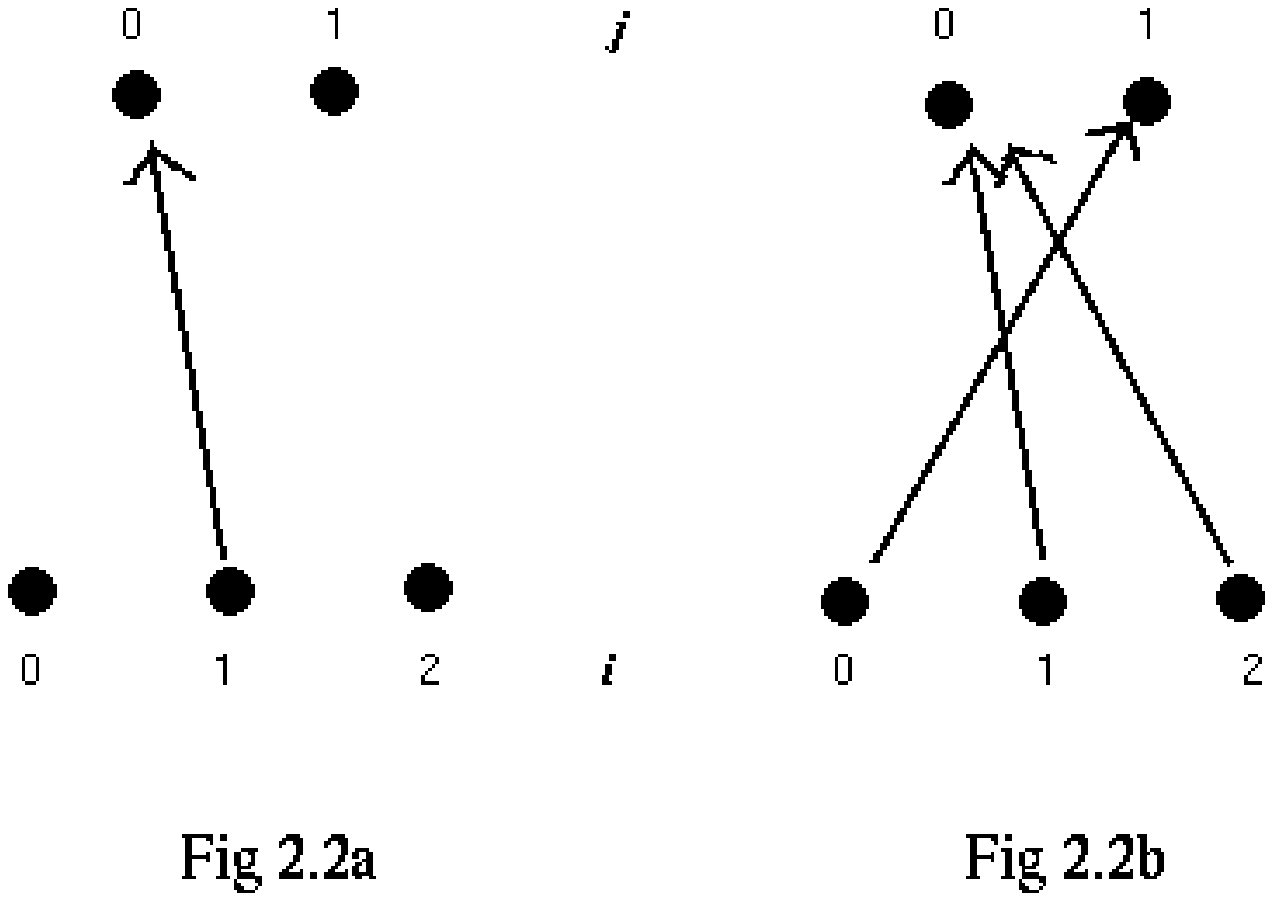} }
\label{fig2.2}

\caption{(a) A thin arrow represents a single transition, in this case from
$i=2$ to $j=1$.  (b)  All transitions for a deterministic system, from each
$i$ to some $j$.  All these transitions together are represented by a
transfer function $F$.
}
\end{figure}
Figures 2.2a,b illustrate the properties of a system $S$ in more
detail, by looking at the transitions between individual values of the
input $i$ and the output $j$.  A single transition from an $i$ to a
$j$ can be represented by an arrow from $i$ to $j$, as in 2.2a for the
transition $2\to 1$ in a system with three input values and two output
values.  For a given deterministic system, there is one transition
for each value of $i$, and all together are given by the transfer
function $F$, which can be represented by the transition arrows from
all the input values $i$, as illustrated in figure 2.2b.  The number
of possible transitions $\N(i\to j)$ between the $i$ and the $j$ is
given by the number of possible arrows pointing from an input value
$i$ to an output value $j$.

Figure 2.3 illustrates all the possible transfer functions for an
elementary binary system.  $F_0$ and $F_1$ are {\it degenerate}, which
means that the input has no effect on the output.  When an output
is degenerate, no signal is sent from the input and there is no
causal connection between them.

\begin{figure}[htb]

\epsfxsize14.0cm
\centerline{\epsfbox{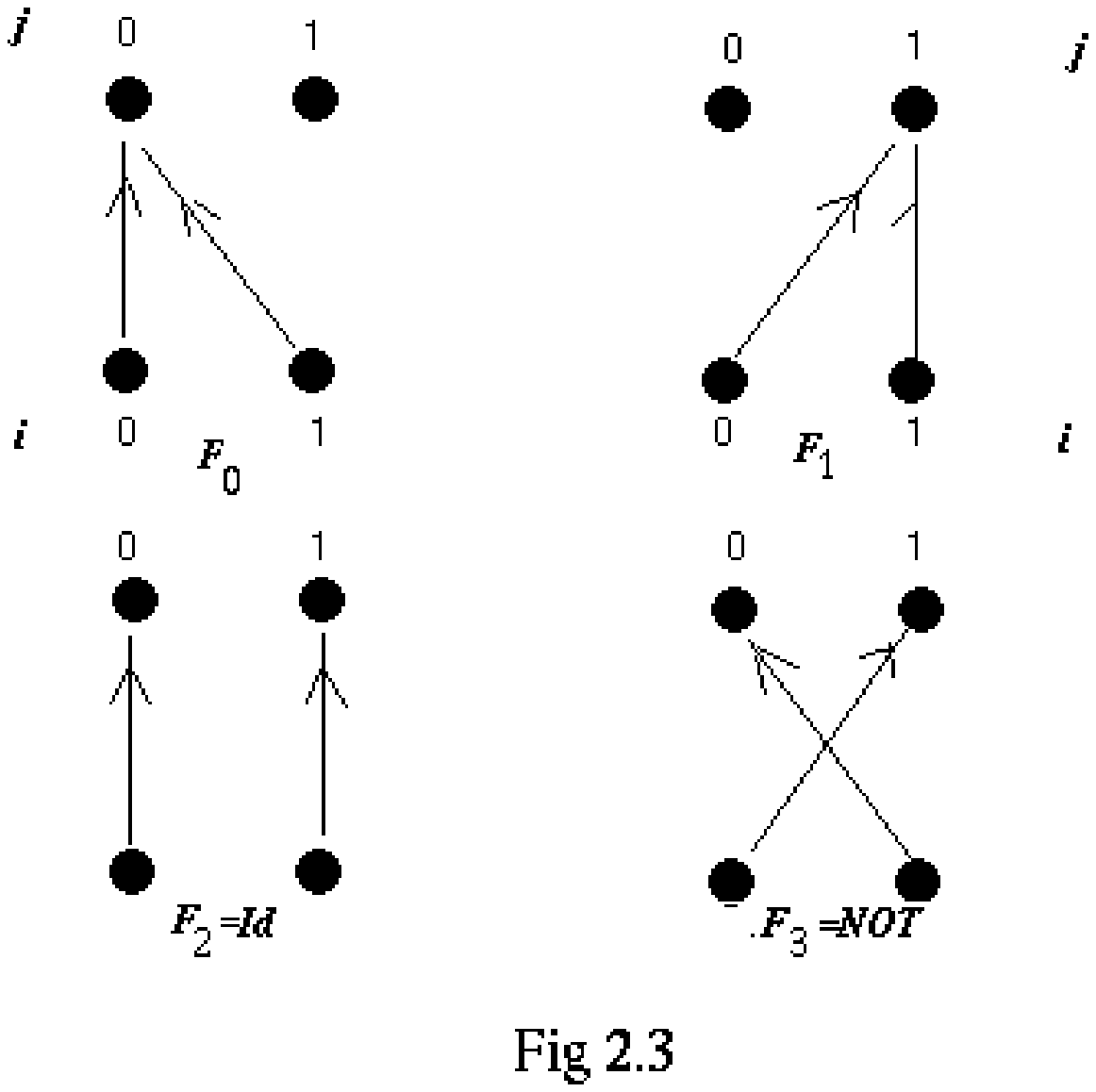} }
\label{fig2.3}

\caption{Transfer functions for an elementary binary system.}
\end{figure}
On the other hand, $F_2$ and $F_3$ are nondegenerate signalling
transfer functions, recognizable as the familiar identity gate $\I$
and the $NOT$ gate.

For the system illustrated in figure 2.2, $\N(i)=3$, $\N(j)=2$.  The
number of possible arrows from an input value to an output value is
just the product of the number of input values and the number of
output values, giving
$$
\N(i\to j)= \N(i)\N(j).
\eqno(2.3)$$
\begin{figure}[htb]

\epsfxsize14.0cm
\centerline{\epsfbox{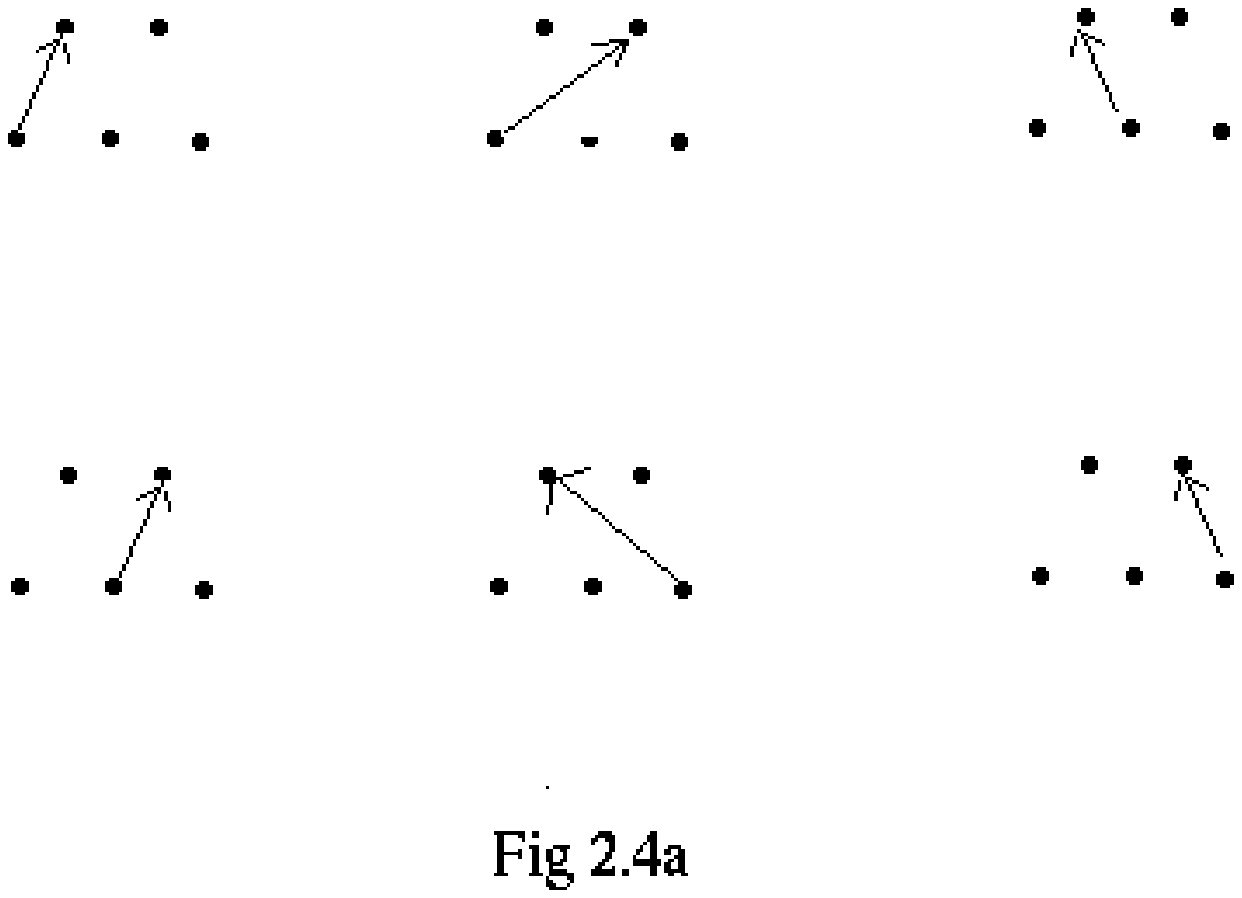} }
\label{fig2.4a}

\caption{The $3\times 2$ transitions $i\to j$ for system 2.2}
\end{figure}
\begin{figure}[htb]

\epsfxsize14.0cm
\centerline{\epsfbox{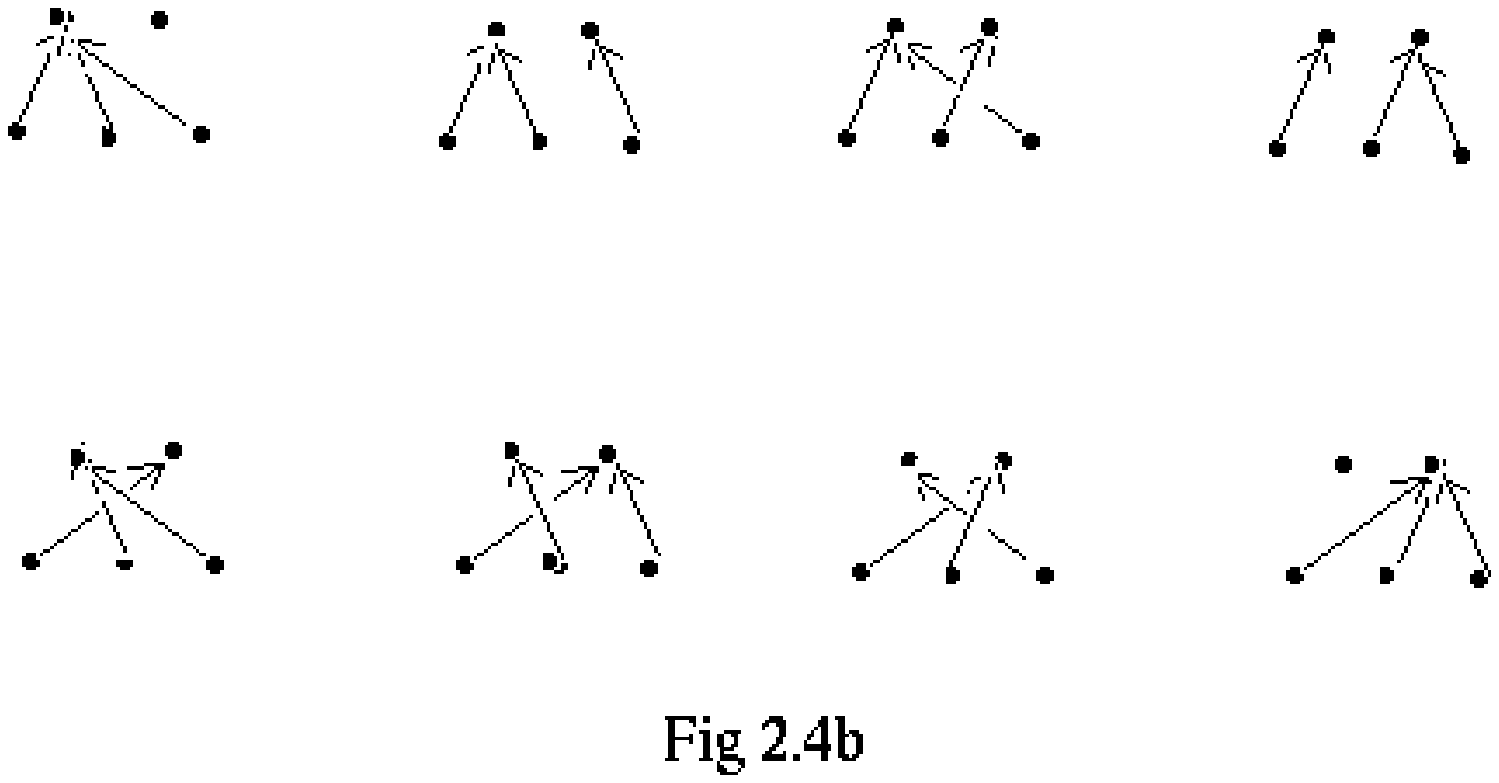} }
\label{fig2.4b}

\caption{The $2^3$ transfer functions $F$ for system 2.2}
\end{figure}
This is illustrated for our system in figure 2.4a.  To define
the dynamics of the system, we need the transfer function $F$, which
corresponds to a set of arrows, one from every $i$ to some $j$.  How
many different possible transfer functions are there?  This is
illustrated for our example in figure 2.4b.  For one $i$ there is a
choice of $\N(j)$ possible arrows $i\to j$, for two $i$'s there is a
choice of $\N(j)^2$ possible pairs of arrows, and for all $\N(i)$
allowed values of the input, there are $\N(j)^{\N(i)}$ possible
choices for the $\N(i)$ arrows, so the number of possible transfer
functions is
$$
\N(F) = \N(j)^{\N(i)}.
\eqno(2.3)$$

The {\it transition picture} of the dynamics is based on the
transitions, and the {\it transfer picture} is based on the transfer
functions.  The distinction between these pictures seems artificial at
this stage, since they are trivially equivalent, but for the
stochastic systems treated later it is crucial.

The number of possible transfer functions is usually much larger than
the number of transitions.  This is important for the theory of
stochastic systems.  We will often have to consider the space of
transfer functions.  This space is discrete for our systems, but
often large in the sense that it consists of a large number of
distinct points, each corresponding to one possible transfer function.

Binary systems, for which each of the inputs $i_p$ and outputs $j_q$
has only two possible values, are particularly simple and important.
The gates which form the elements of digital circuits computers are
examples, including the NOT, AND and control-NOT gates, together with
the elementary binary system for which the transfer function is the
identity.

A general deterministic system $S$ with two input and two
output ports has a transfer function $F$ for which
$$ 
(j_1,j_2) = F(i_1,i_2).
\eqno(2.4)$$
Each output $j_q$ is a function $F_{j_q}(i_1,i_2)$ of the inputs, so $F$ can
be expressed as a `vector' 
$$ 
F = (F_1,F_2),\hx{where}\h10 j_1 = F_1(i_1,i_2),\h10 j_2 = F_2(i_1,i_2).
\eqno(2.5)$$ 
The causal links between the ports are illustrated in Figure 2.5a.
\begin{figure}[htb]

\epsfxsize14.0cm
\centerline{\epsfbox{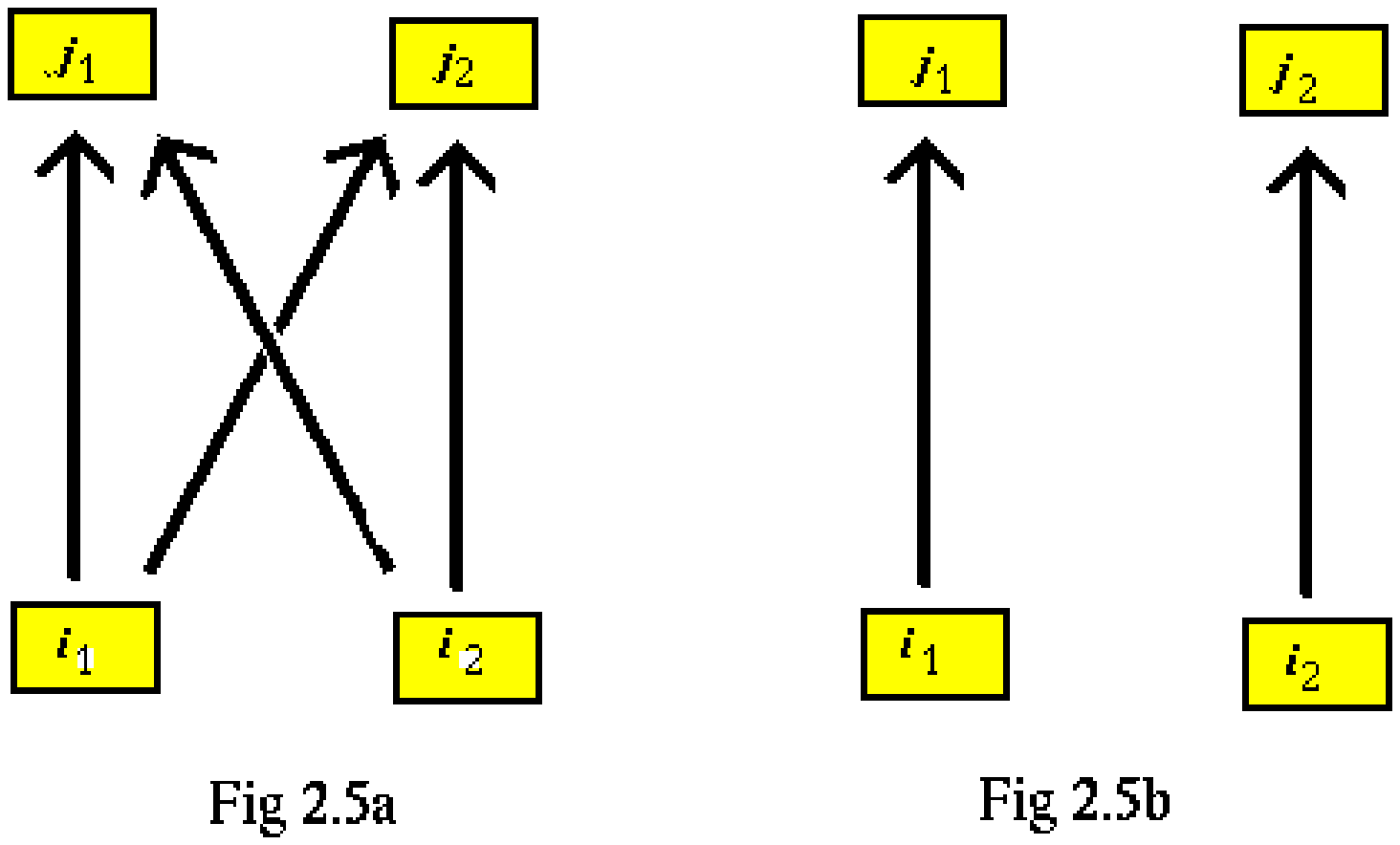} }
\label{fig2.5}

\caption{Causal links, (a) for a general system with two input and two output ports
and (b) for the specific case when it is formed of two independent
elementary subsystems}
\end{figure}
Suppose all four ports of the two subsystems are binary, then the
total number of inputs or outputs is 4.  There are $4\times 4=16$
possible transitions, and the total number of transfer functions is
$4^4=256$.

\vs{\bf 2.3 Combining systems}

Systems $S^k$ can be combined together in various ways to make a new
system $S$, so that the $S^k$ become subsystems of $S$.  We make the
standard assumption that the causal relation between an input and an
output of a subsystem $S^k$ is unaffected by this process.  The
external relations of a subsystem do not affect its internal dynamics.
The transitions and the transfer functions are therefore unaffected.

Properties of different systems, such as transfer functions, are
usually distinguished by an index, to distinguish them from properties of
the same system, distinguished by a suffix.  When we combine systems,
the notation can become ambiguous, and either may be used.

Suppose there are two elementary systems $S^1$ and $S^2$,
each with a single input and a single output port, with inputs and
outputs $i^1,i^2$ and $j^1,j^2$ and transfer functions $F^1$ and
$F^2$.  If these subsystems are independent, then they can be
considered together as a single system $S$, which is not elementary
because it has two inputs $i_1,i_2$ and two outputs $j_1,j_2$.  These
may however, be considered together to form {\it the} input
$i=(i_1,i_2)$ and {\it the} output $j=(j_1,j_2)$ of the system
$S$, as illustrated in figure 2.6.  
\begin{figure}[htb]

\epsfxsize14.0cm
\centerline{\epsfbox{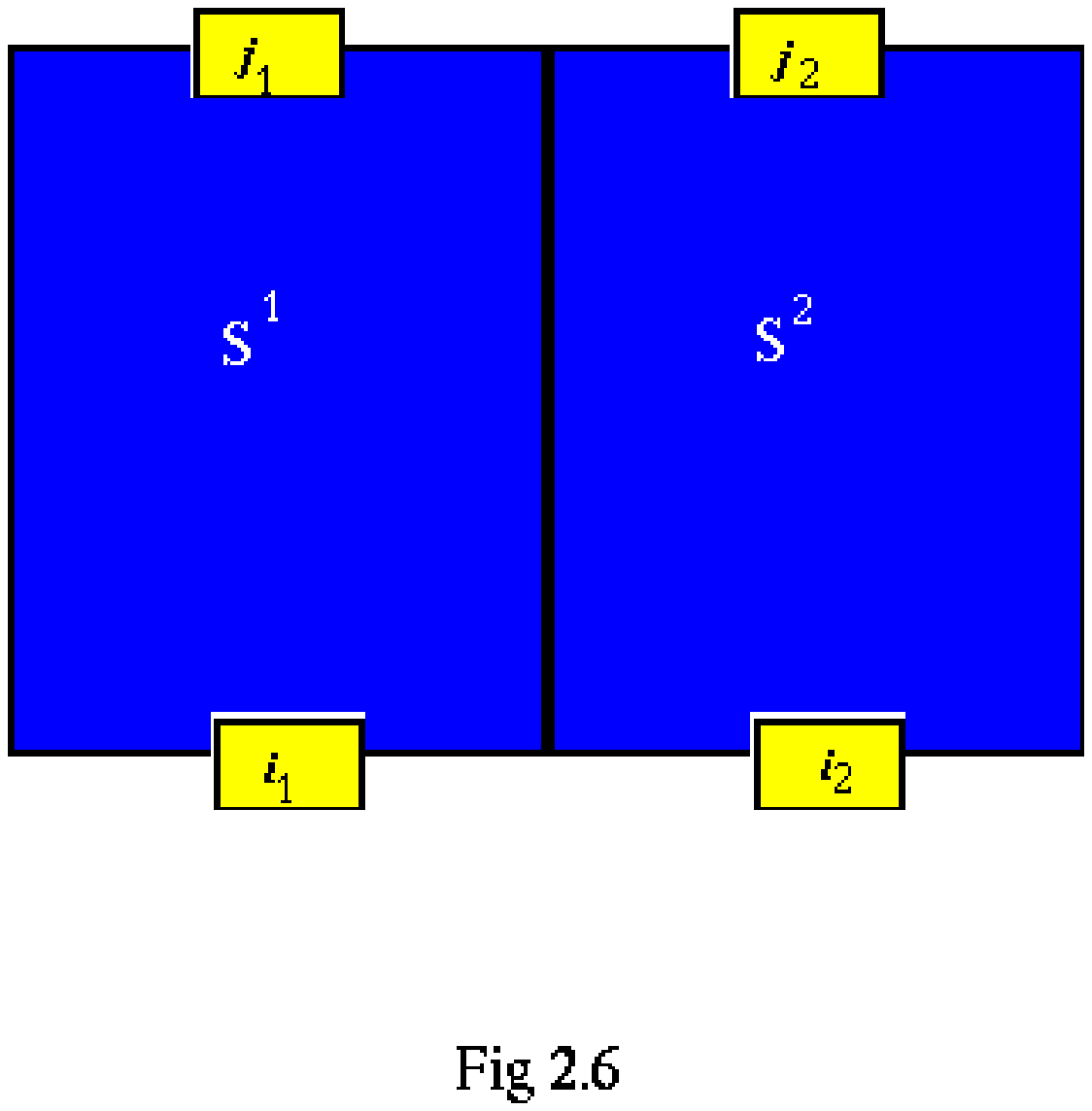} }
\label{fig2.6}

\caption{Combination of independent systems $S^1$ and $S^2$ to form
a system $S$ with $i=(i_1,i_2)$, $j=(j_1,j_2)$}
\end{figure}
When a system $S$ is made from a combination of two separate and
independent subsystems $S^1,S^2$ in this way, the transfer functions
$F_j=F^j$ have the special form
$$
j_1=F^1(i_1,i_2)=F^1(i_1),\h10  j_2=F^2(i_1,i_2) = F^2(i_2).
\eqno(2.6)$$
The port $j_2$ is not affected by $i_1$, and nor is $j_1$ affected by
$i_2$.  There are no signals between the subsystems, as
illustrated in figure 2.5b.  The total number of transitions
and transfer functions is
$$
N(i\to j) = N(i_1\to j_1) + N(i_2\to j_2)\hx{and}\h10
N(F) = N(F^1)N(F^2).
\eqno(2.7)$$
When all four ports are binary, then there are $2\times 2+2\times 2=8$
transitions and $2^2\times 2^2=16$ transfer functions.  When compared
with the general system with four binary ports described at the end of
the previous section, the combined system has fewer transitions and
far fewer transfer functions.

\begin{figure}[htb]

\epsfxsize14.0cm
\centerline{\epsfbox{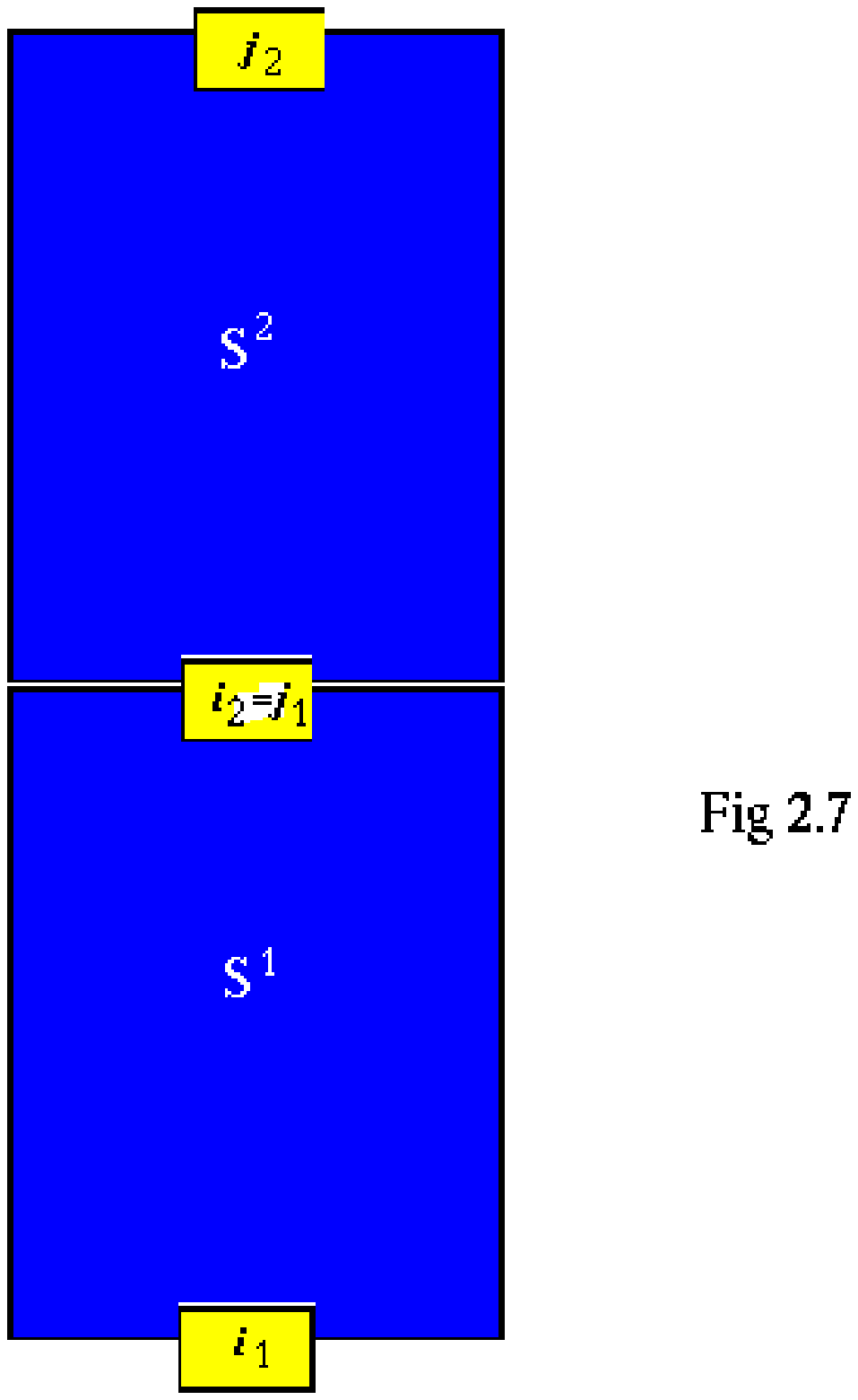} }
\label{fig2.7}

\caption{Systems $S^1$ and $S^2$ combined in series to form a system 
$S$.}
\end{figure}
Now suppose our two elementary systems are joined {\it in series} so
that the output port of $S^1$ is the input port of $S^2$.  As a black
box, the combined system has only one input $i=i_1$ and one output
$j=j_2$, which are illustrated in figure 2.7.  Joining the
systems has reduced the number of ports by one input and one output
port, and
$$
j = F^2(F^1(i)) = F(i),\hx{where}\h10 F=F^2\o F^1, 
\eqno(2.8)$$ 
and the standard $\o$ notation has been used for the composition of
the transfer functions.  The same function $F$ may result from
different combinations of $F^1$ and $F^2$.  The number of transfer
functions is given by equation (2.3), with $i=i_1, j=j_2$.  Evidently
there is a consistency condition for these two systems to be combined
in series, that the output port $j_1$ must be of the same type as the
input port $i_2$, and in particular that $\N(j_1)=\N(i_2)$.

The generalization of the theory for two elementary systems to an
arbitrary number of elementary systems is clear.  For elementary
systems in series, the same transition can result from many different paths 
through the input and output ports of the the subsystems,
and the same transfer function from many different combinations
of transfer functions for the subsystems.

The generalization to many systems is also clear for {\it independent}
systems that are not elementary.  But it is certainly {\it not} true
for arbitrary combinations of systems which are joined by different
input and outputs ports $i_p$ and $j_q$ as illustrated for some
relatively simple cases in figures 2.8 and 2.9.  Such arbitrary
combinations include every digital communication system and every
digital computer that has been built!
\begin{figure}[htb]

\epsfxsize14.0cm
\centerline{\epsfbox{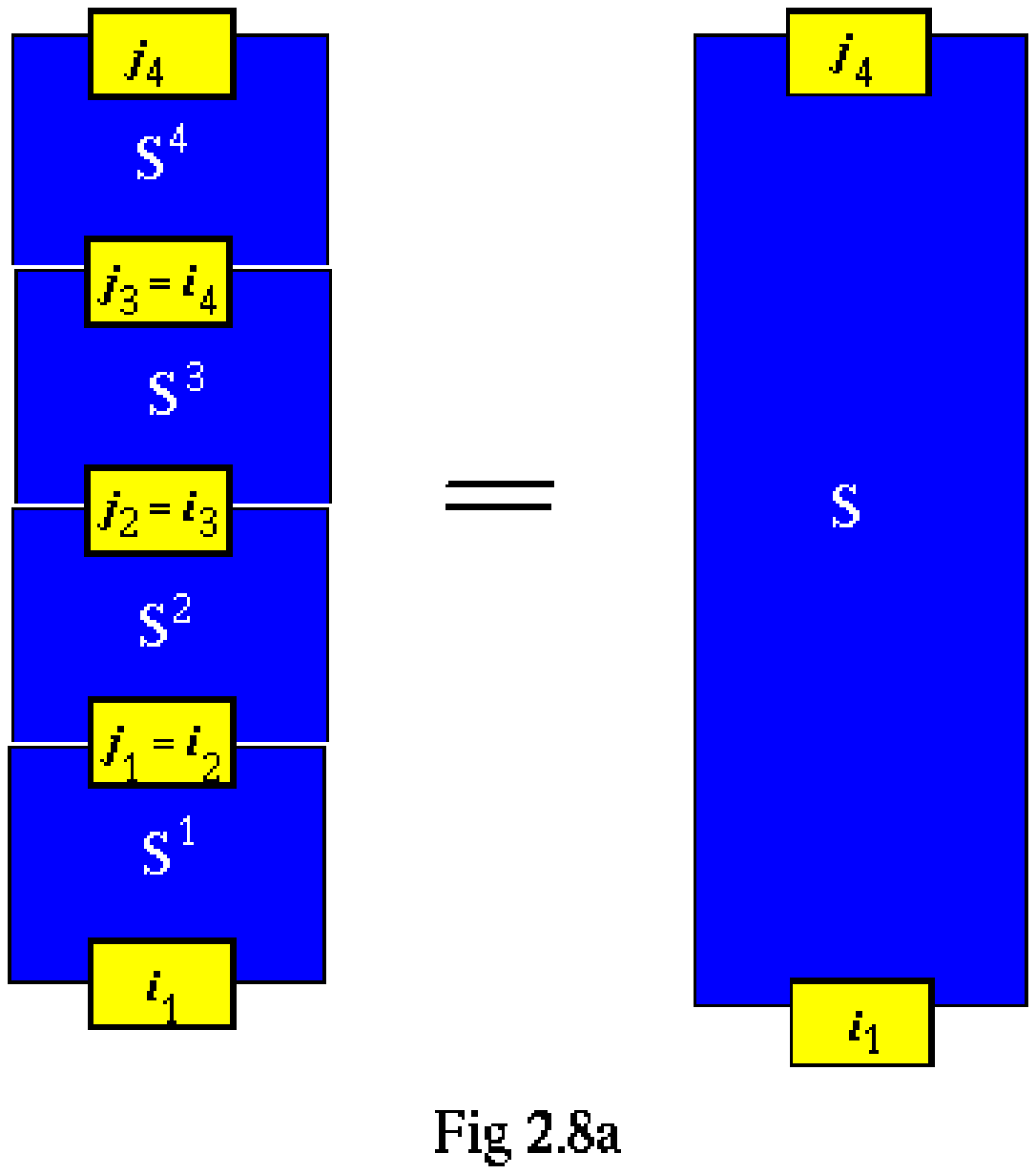} }
\label{fig2.8a}

\caption{A chain of four subsystems connected in series.}
\end{figure} 
\begin{figure}[htb]

\epsfxsize14.0cm
\centerline{\epsfbox{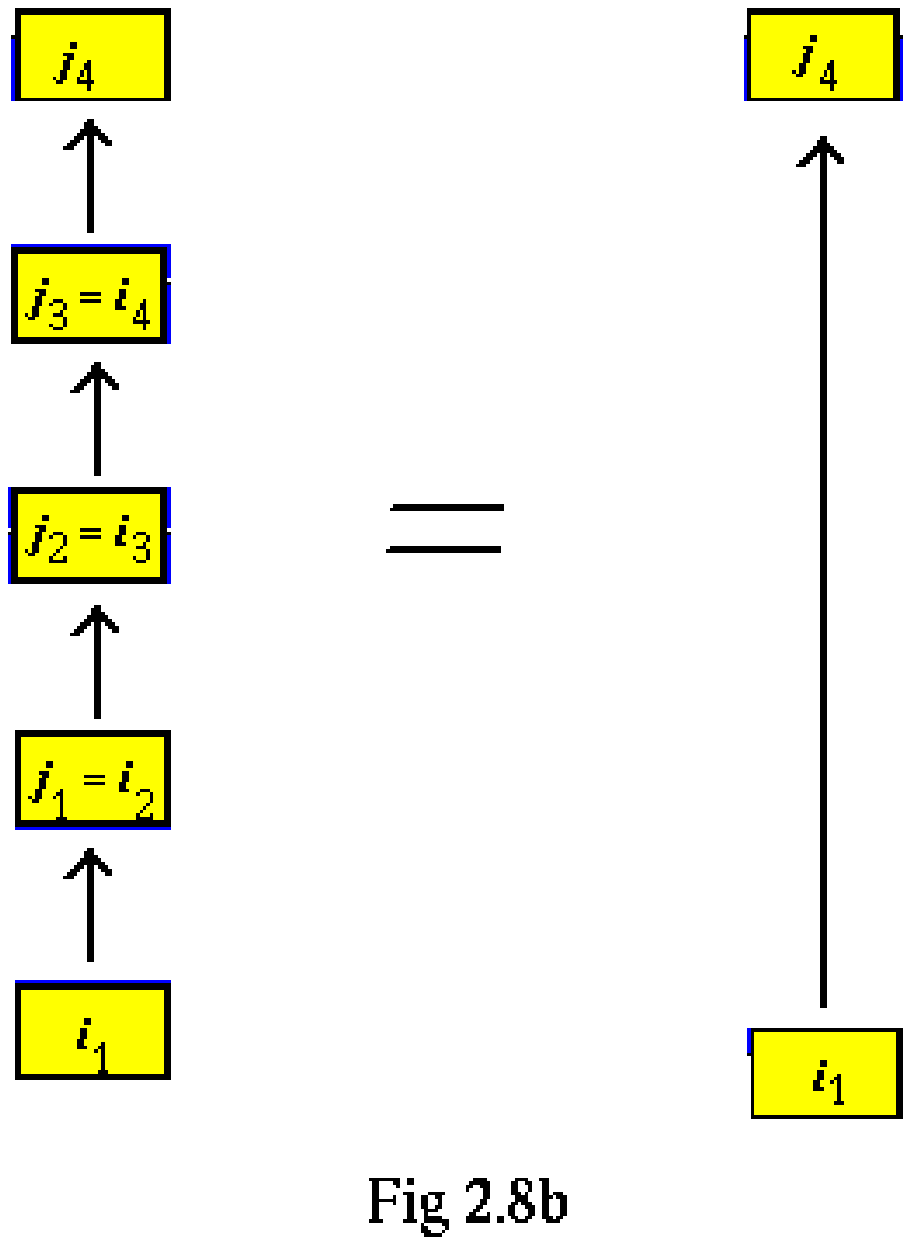} }
\label{fig2.8b}

\caption{A chain of four subsystems connected in series: causal connections.}
\end{figure} 
\begin{figure}[htb]

\epsfxsize14.0cm
\centerline{\epsfbox{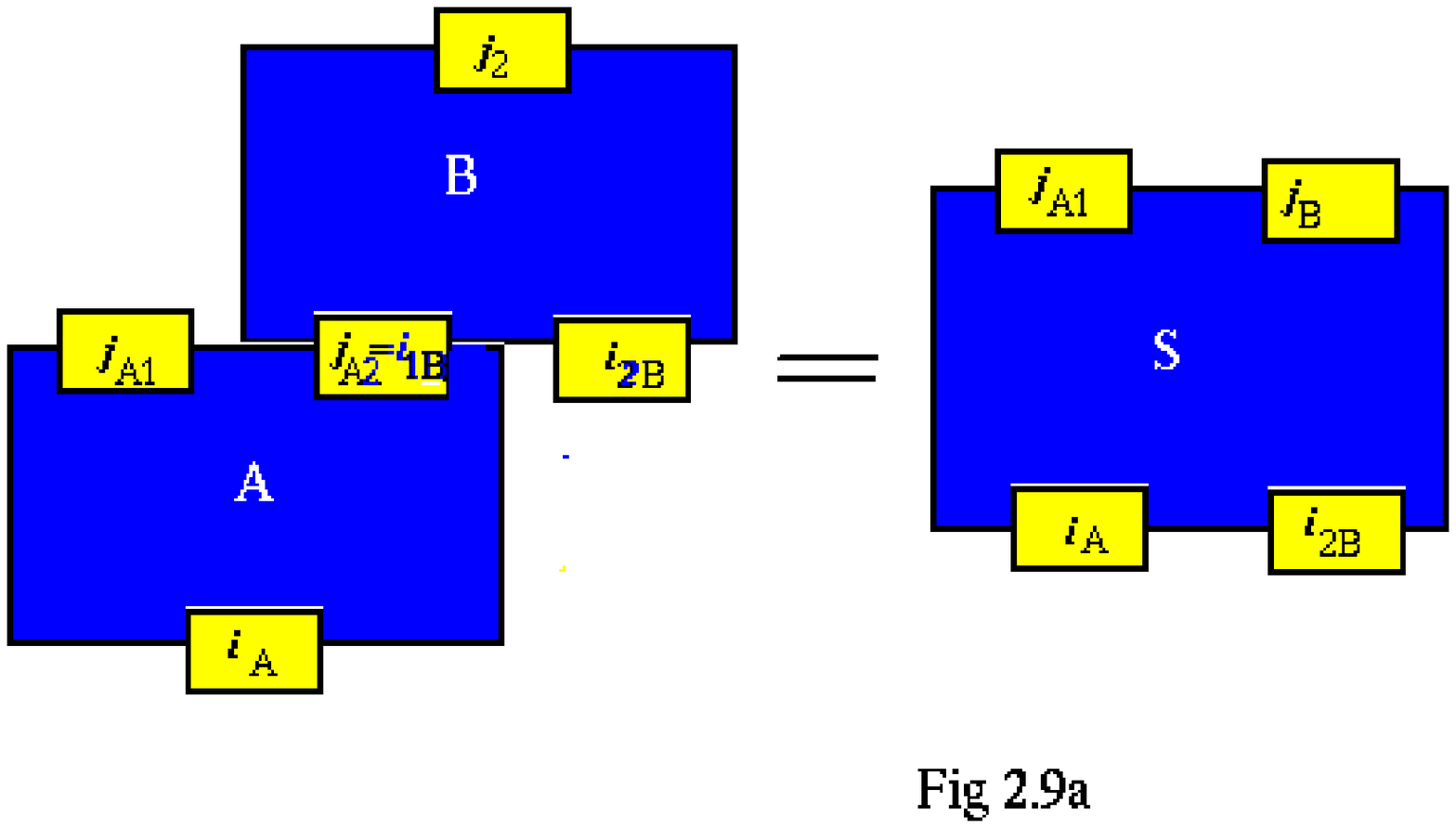} }
\label{fig2.9a}

\caption{Systems joined by less than the maximum number of ports.}
\end{figure} 
\begin{figure}[htb]

\epsfxsize14.0cm
\centerline{\epsfbox{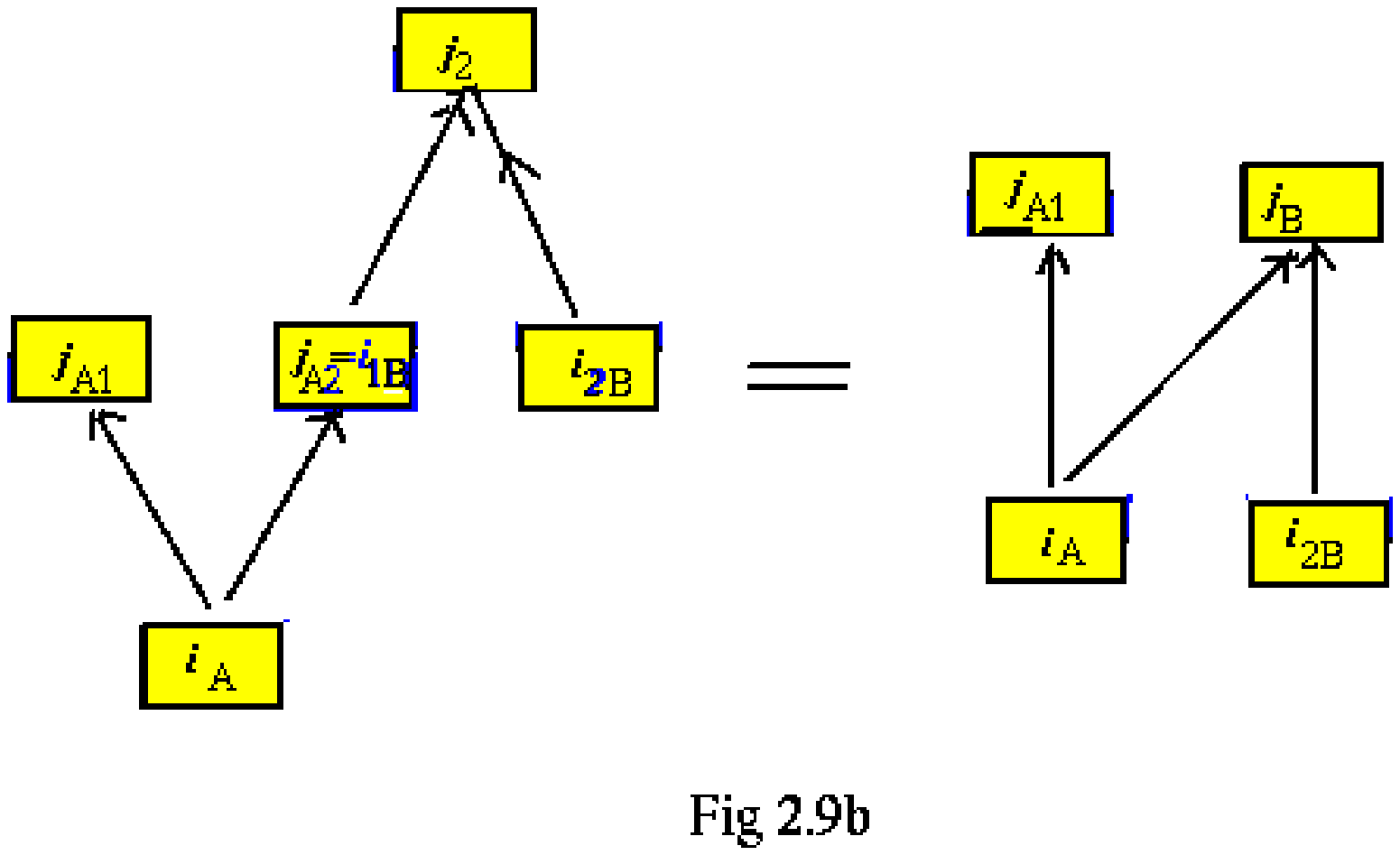} }
\label{fig2.9b}

\caption{Systems joined by less than the maximum number of ports: 
causal connections.}
\end{figure} 
We have seen that a system formed by the combination of the
two independent systems has fewer possible transitions than for a
general system with the same ports, and the the number of transfer
functions for the combined system can be considerably less.

There are so many different kinds of linkages like this, that we will
consider them independently as they arise.

\vs{\bf 2.4 Causal loops}

Some of the following sections present the input and output in the
context of space and time, which puts constraints on the possible ways
of linking systems through their inputs and outputs, but in this
section we ignore such constraints.  Most of us believe it is not
possible to form a causal loops because of their time relations,
although some cosmologists who work with wormholes might disagree.
Effects are expected to come after causes, outputs after inputs, so
the final output from a chain of systems like that in figure 2.8
cannot be the same as the initial input.  However, in relativity, time
cannot be considered without space, and the spacetime relations of
systems which include quantum measurements are particularly
subtle.  

We need to consider causal loops in the abstract first, to distinguish
spacetime constraints from those that are independent of spacetime.
We now demonstrate that there are such independent constraints.

We make the following provisional assumptions about deterministic
systems:

AD1.  A finite number $\N(S^k)$ of subsystems $S^k$, with fixed
input and output ports and transfer functions $F^k$ can be
combined in any way to form a system $S$, provided that the
corresponding input and output ports are consistent.

AD2.  Combining systems does not change their transfer functions $F^k$.

For some collections of subsystems, these assumptions lead to a
contradiction, as we now show.

\begin{figure}[htb]

\epsfxsize14.0cm
\centerline{\epsfbox{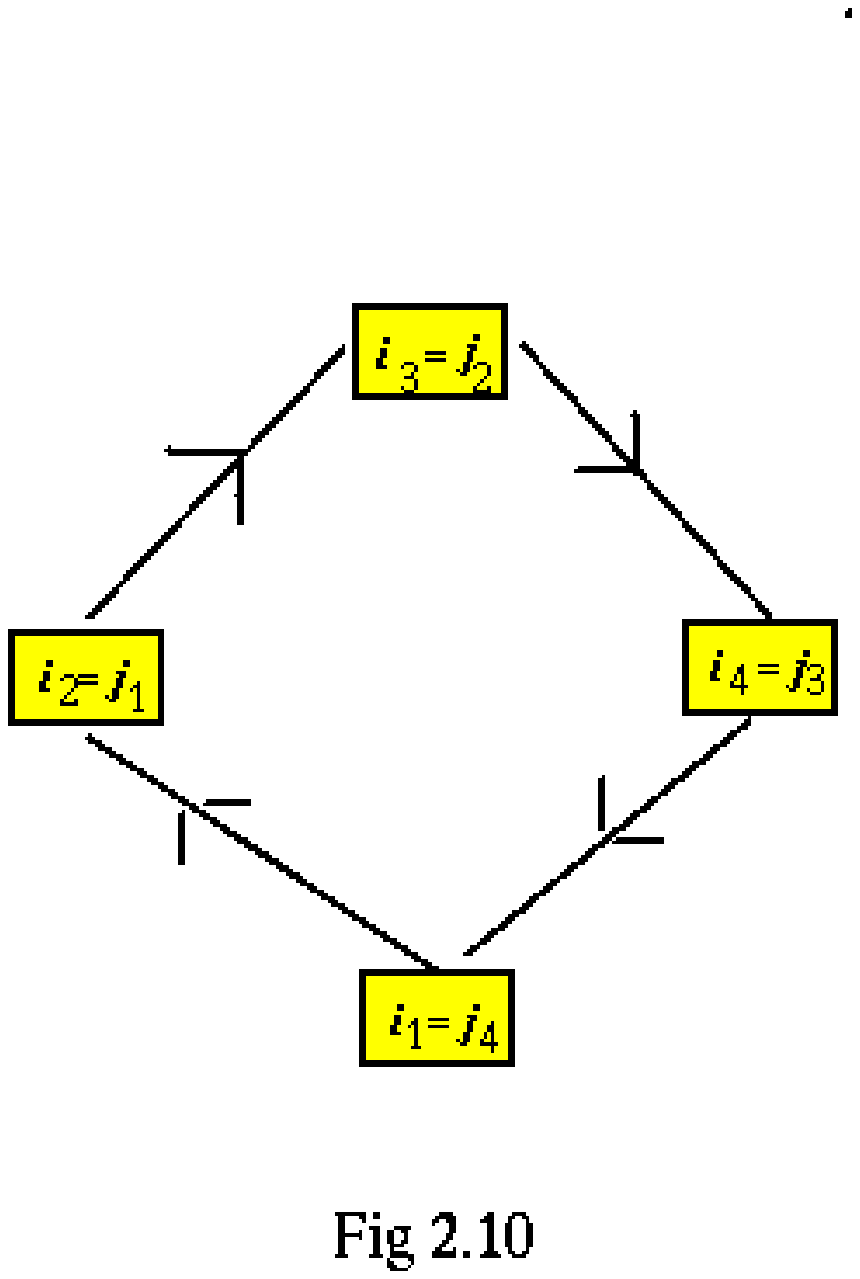} }
\label{fig2.10}

\caption{A causal loop formed from four subsystems.}
\end{figure} 
Two or more subsystems may be linked together in a loop, which we shall
call a causal loop, as illustrated in figure 2.10 for four systems.

Causal loops in spacetime are not the same as feedback loops in space.
The output of a feedback loop is alway delayed with respect to the 
input, it comes to the same point in space, but later in time, whereas
for a causal loop the output and input are the same, at one point
in spacetime.  A feedback loop may be unstable, whereas
a causal loop can be forbidden.
 
In general, each system $S^k$ with $k=1,\dots \N(S)$ has input $i^k $
and output $j^k=i^{k+1}$, which is also the input to the next system.
The corresponding transfer function is $F$, where $i^{k+1}=F^k(i^k)$.
The loop is closed by $j^{\N(S)}=i^1$, from which it follows that
$$
i^1 = F_{\rm loop}(i^1),\hx{where}\h10 F_{\rm loop} = F^{\N(S)}
\o\dots\o F^2\o F^1. 
\eqno(2.9)$$ 

Suppose, however, that in the case of the four subsystems illustrated
in figure 2.10, all the ports are binary and the systems have 
transfer functions
$$
F^1 = F^2 = F^3 = \I, \h10 F^4 = NOT, 
\eqno(2.10)
$$ 
so the loop transfer function is $F_{\rm loop} = NOT$.  Then from
equation (2.9), $i^1=0$ implies $i^1=1$ and $i^1=1$ implies $i^1=0$,
so there is a contradiction.  This loop transfer function is
forbidden.  One of the assumptions must be wrong.

In general, a loop transfer function $F_{\rm loop}$ must have at least
one input with an identical output if it is to be allowed.  The inputs
that satisfy this condition are the allowed inputs.  If there is such
an input, then the causal loop is allowed.  If there is no such input,
then the causal loop always leads to a contradiction.  This
combination is forbidden, so for any collection of subsystems which
can be combined into a forbidden causal loop, one of the assumptions
must be wrong.

This is the {\it deterministic loop constraint}.

This constraint does not depend on spacetime relations, but they are 
connected.  For example, in classical nonrelativistic
dynamics, effects always come after causes, so the assumption AD1 is
always wrong: it is not possible to construct any causal loops, and the
distinction between allowed and forbidden causal loops is irrelevant.
So once the spacetime constraints are used, the loop constraint tells
us nothing more.  The same is true in classical special relativity.

This loop constraint has applications in classical general relativity,
in which there are solutions of Einstein's equations with `wormholes'.
The problems of causality for these curved spacetimes with 
unusual topologies are sometimes discussed in terms of such violent
acts as `shooting your grandmother before she conceives your father'.
The deterministic loop constraint is a formal statement of this and
also more gentle contradictions.  However, these applications are not
our main concern, for the deterministic loop constraint is used here
as an introduction to the stochastic loop constraint of section 3.4,
which can be applied quantum measurement.

\vs\vs\cl{\bf 3 Stochastic systems}

\vs{\bf 3.1 Transition and transfer pictures}

The theory so far has been restricted to deterministic systems.  A
system is deterministic if the value $j$ of the output is uniquely
determined by the value of $i$ of the input.  The dynamics of a
deterministic system can be described in two ways that are trivially
equivalent.  The first is to give the output $j$ for each input $i$.
This is the transition description.  The other is to give the transfer
function $F$ that defines this relation between outputs and inputs:
$j=F(i)$.  This is the transfer description.

More generally, classical systems are stochastic, or noisy.  According
to classical mechanics this is the result of interactions with other
systems that cannot be eliminated.  These other systems will be called
the {\it environment} of the original system.  For a system $S$, the
distinction between another system $S'$ of the type considered here
and a part of the environment is a matter of convention, as it is in
any other field in which there are systems and an environment.  An
example of a noisy system is a noisy NOT gate, in which, very
occasionally, an input $i=0$ results in an output $j=0$ instead of the
wanted $j=1$.  In practice such noise cannot be avoided with certainty
in digital circuits, although its probability can be made very low.

The results of quantum measurements are also typically stochastic, and
according to ordinary quantum theory, this departure from determinism
does not depend on the environment.  Here we do not consider quantum
systems on their own.  A system always has inputs and outputs that are
classical.  In a quantum measurement, the inputs are the values of the
macroscopic classical variables that determine which dynamical
variables of the quantum system are being measured, and the outputs
are the macroscopic classical states of the measuring and recording
equipment that are produced by the measurement.  So the system
consists of the quantum system being measured, together with the
essential classical parts of the apparatus being used to measure it.
\begin{figure}[htb]

\epsfxsize14.0cm
\centerline{\epsfbox{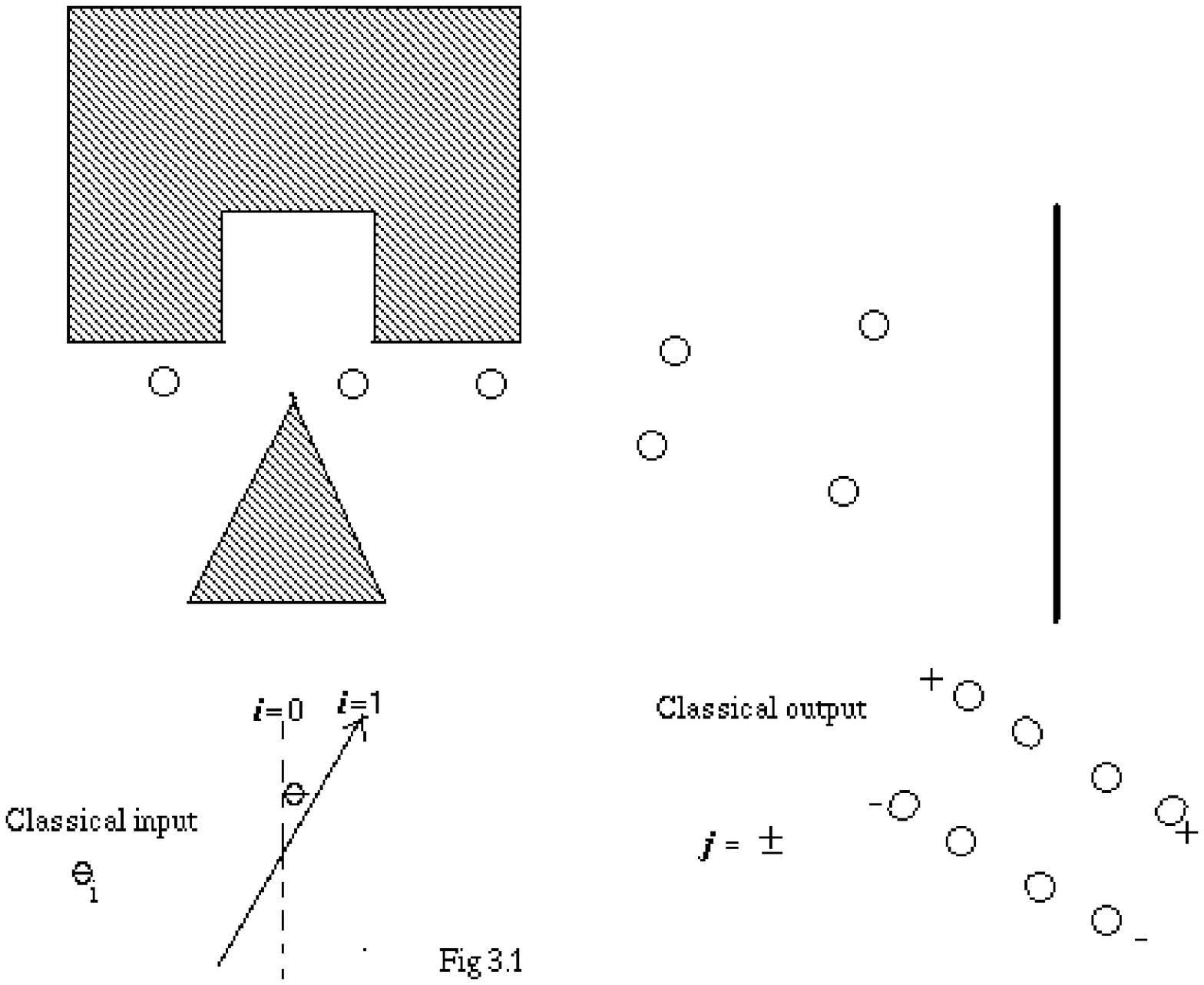} }
\label{fig3.1}

\caption{Stern-Gerlach experiment with variable direction of the 
magnetic field.}
\end{figure} 
For a Stern-Gerlach experiment with individual atoms, the input is the
orientation of the magnetic field, and the output is the classical
record of the direction of spin with respect to the field, as
illustrated in figure 3.1.  We assume that the spin of the incoming
atoms is vertically upwards.  The orientation of the magnetic field
determines which dynamical variable is being measured.  According to
Bohr, the result depends on the conditions under which the measurement
is made \cite{Bohr1935,Bohr1983}.  In some modern experiments, such as
those to test nonlocality through Bell's inequality, these conditions
must be set or changed during the course of an experiment, and these
are the inputs.  Bohr did not mention this possibility.

For an experiment of the Bell type, the inputs are the orientations
of the polarizers, and the outputs are the classical records of the
measured polarization.  In accordance with our insistence on discrete
inputs, we consider only a finite number of possible orientations of
the magnetic field for the Stern-Gerlach experiment or of the
polarizers for the Bell experiment.  The outputs are discrete because
the spin components are quantized.

Just as for deterministic systems, the dynamics of stochastic systems
can be described in terms of transitions $i\to j$ or in terms of
transfer functions $F$, but for stochastic systems the relation
between them is not trivial.  In the transition picture the dynamics
is defined by the probabilities of the transitions
$$
\Pr(i\to j) = \Pr(j|i),
\eqno(3.1)$$
that is, the conditional probability of the output $j$ given the input
$i$.  Because there is always {\it some} output, these sum to unity
for every $i$:
$$
\sum_j\Pr(j|i)=1.
\eqno(3.2)$$

In the transfer picture, the dynamics is defined by probability that
the system behaves like a deterministic system with transfer function
$F$, for every possible $F$.  These are the transfer function
probabilities, or transfer probabilities, $\Pr(F)$, which sum to unity.

The conditional probability for finding the output $j$, when the input
is $i$, is then given in terms of the transfer probabilities as
$$ 
\Pr(j|i) = \sum_F\Pr(F)\delta(j,F(i)).
\eqno(3.3)$$

This gives the transition probabilities in terms of the transfer
probabilities.  What about the other way round?  There are nearly
always many more transfer functions than transitions, the space of
transfer probabilities is of higher dimension than the space of
transition probabilities and so for a given set of transition
probabilities $\Pr(j|i)$, there cannot in general be a unique set of
transfer probabilities $\Pr(F)$.  Deterministic systems, for which all
probabilities are zero or one, are an important exception.  Given the
transition probabilities $\Pr(j|i)$, the transfer probabilities
$\Pr(F)$ must satisfy equation (3.3), together with the inequalities
and normalization of the transfer probabilities,
which are
$$
0\le\Pr(F)\le 1,\h10 \sum_F \Pr(F)=1,
\eqno(3.4)$$
but (3.3) and (3.4) do not define the transfer probabilities uniquely.

We show later that the inequalities in (3.4) include the Bell
inequalities.

Since there are usually more possible transfer functions than
transitions, fixing the transition probabilities does not fix the
transfer probabilities.  For a system with given input and output
ports, and without additional constraints, the number of independent
transition probabilities is
$$
\N(\Pr(i\to j)) =\N(i)(\N(j)-1), \eqno(3.5)$$ where the $-1$ comes
from the condition that the transition probabilities from a given $i$
must sum to unity.  The number of independent transfer probabilities
$\Pr(F_k)$ is $\N(F)-1$, which is usually greater.  Within the space
of sets of transfer probabilities, one for each $F_k$, there is a
region of dimension given by equation (3.5) which is consistent with
given transition probabilities.  This we will call the {\it consistent
region} of transfer probabilities, although strictly speaking it is a
consistent region of sets of transfer probabilities, where a set
contains a probability $\Pr(F_k)$ for each transfer function $F_k$.
This consistent region is the transfer picture equivalent of the 
set of transfer probabilities $\Pr(i\to j)$.

Frequently there {\it are} additional constraints, and they are often
relatively simple to express in the transfer picture.  The constraints
may come from the physics, or they may be imposed in order to test a
hypothesis, such as locality or relativistic invariance.  In such
cases the consistent region of transfer probabilities is reduced,
sometimes to an empty region, in which case the constraints are
incompatible, so that at least one of them must be wrong.

The causal relations for a noisy or stochastic system can be expressed
in terms of signals just as they were for a deterministic system.  If
the probability $\Pr(F)$ of a signalling transfer function is always 
zero, then there are no corresponding signals for the noisy system.
However, because the transfer probabilities are not unique, some
care is needed in applying this principle.

Consider the elementary binary system of figure 2.3, with
the degenerate transfer functions
$$
F_0(i) = 0,\h10 F_1(i)=1 
\eqno(3.5)$$
and the signalling transfer functions
$$
F_2(i)=i,\hx{and}\h10 F_3(0) = 1,\hs\hs F_3(1)=0.
\eqno(3.6)$$

Suppose that the output $j$ is like the unbiassed toss of a coin, so
that it is independent of the input, there is no signal or causal
link from input to output, and
$$
\Pr(j=0)=\Pr(j=1)=\half.
\eqno(3.7)$$
The transition probabilities are then
$$
\Pr(0|0)=\Pr(0|1)=\Pr(1|0)=\Pr(1|1)=\half,
\eqno(3.8)$$
which are unique.  

But the transfer probabilities $\Pr(F)$ are not unique.  One
possibility is
$$
\Pr(F_0) = \Pr(F_1)=\half,
\eqno(3.9)$$
in which the probability of the signalling transfer functions is
zero.  But another is
$$
\Pr(F_2) = \Pr(F_3)=\half, \eqno(3.10)$$ in which the probability of
both signalling transfer functions is non-zero, even though there is
no signal.  The two sets of transfer probabilities (3.9) and (3.10)
give the same set of transition probabilities (3.8).  So for a
stochastic system with a given set of transition probabilities,
signalling transfer functions can have nonzero probability, despite
there being no signal.  A nonzero signalling transfer function
probability does not necessarily indicate that there is a signal.

However, if there is a signal, then for {\it every} set of
probabilities $\Pr(F)$ in the consistent region, at least one
signalling transfer function has nonzero probability.  Clearly this
condition is not satisfied in the above example, but it is satisfied
for every stochastic system with a signal.

\vs{\bf 3.2 Background variables}

For the classical dynamics of an electrical circuit with resistors,
those internal freedoms of the resistors which produce noise in the
circuit are background variables, which are hidden, because we cannot
see the motion of the `classical electrons'.  Unlike some background
variables of quantum mechanics, there is no physical principle that
prevents the hidden variables of the classical circuit from becoming
visible.
 
The same analysis can be applied to a quantum system prepared in a pure
state, with a very important difference: the background variables are
now hidden variables in principle and they need not have the properties
of classical background variables.  When they do not, the dependence
of the output on the input cannot be classical: it must have a quantum
component.  The dependence of the outputs on the inputs cannot then be
produced by any classical system.  A quantum system can do things that
a classical system cannot, which is the basis of technologies like
quantum cryptography and quantum computation.  If the system is in a
black box, an experimenter can sometimes tell, just by controlling the
input and looking at the output, that the output is linked to the
input by a quantum system.  This is why the study of hidden variables
in quantum mechanics has had such important physical and practical
consequences.

The variable $\lambda$ is a complete background variable of a system
if the input $i$ and the background variable $\lambda$ together
determine the output $j$.  Another way of saying the same thing is
that $\lambda$ determines the transfer function, which we then write
as $F_\lambda$, so that
$$
j= F_\lambda(i).
\eqno(3.11)$$
So the output $j$ depends only on the background variable $\lambda$
and the input $i$, and is uniquely determined by them.  There are
incomplete background variables for which the output is not uniquely
determined, but we will suppose that they are supplemented if
necessary by further background variables to make the output unique
for a given $i$ and $\lambda$.  From now on, all background variables
are complete.

Given the probability $\Pr(\lambda)$ for the background variable, the
probability of the output $j$, given the input $i$, is
$$
\Pr(j|i) = \int\d\lambda\hs \Pr(\lambda)\delta(j,F_\lambda(i)).
\eqno(3.12)$$
For a resistor, $\Pr(\lambda)$ is just the appropriate Boltzmann
distribution.

Transfer function analysis is particularly suitable for background
variables.  For a particular system, two values of the background
variable $\lambda$ which have the same transfer function $F_\lambda$
are completely equivalent.  For a given input they result in the same
output.  Only the function $F_\lambda$ matters.  So all the
information that is needed about the probability distribution
$\Pr(\lambda)$ is contained in the finite number of transfer
probabilities for the transfer functions, which are
$$ 
\Pr(F) = \int\d\lambda\hs\Pr(\lambda)\delta(F,F_\lambda).  
\eqno(3.13)$$
Once we have the transfer probabilities for a system $S$, we can
forget about the probability distribution for the background variables
of $S$.  This is a great help, because the transfer probabilities
usually live in a much simpler space.

We have seen that in general, systems with given transition
probabilities do not have unique transfer probabilities.  But we now
see that systems with complete background variables {\it do} have
unique transfer probabilities $\Pr(F)$, which therefore contain
information about the relation of the system to its environment
which is not contained in the transition probabilities.  Also, for a
given system, the transfer functions themselves can be treated as if
they were discrete background variables.

The conditional probability for finding the output $j$, when the input
is $i$, is then given by the sum
$$ 
\Pr(j|i) = \sum_F\Pr(F)\delta(j,F(i)),
\eqno(3.14)$$
which is usually much simpler in practice than the integral (3.12)
over the background variable $\lambda$.  This links the theory of
background variables, including the possible hidden variables of
quantum mechanics, with transfer functions.  

But we should remember that the region of transfer probabilities which
is consistent with the transition probabilities, is a general property
of stochastic input-output systems, defined in terms of the
transition probabilities, and independent of any hidden or background
variables that there may or may not be.

\vs{\bf 3.3 Independence}

Independence of systems is more subtle for stochastic systems than it
is for deterministic systems.  Subsystems of a classical stochastic
system $S$ may be correlated by connecting inputs and outputs.  But
they may also be correlated through the environment, which is not part
of $S$.  Quantum systems produce their own very special kind of
correlation, in which classical outputs corresponding to two separate
and spatially separated measurements of the same quantum system can be
correlated as a result of quantum entanglement.  This is particularly
important for Bell-type experiments.

These properties can be expressed in terms of background variables, or
entirely in terms of the transfer functions, without the need for
background variables.  Now consider some special cases.

Two systems that are completely independent, without any correlation
through their background variables, or through entanglement, have
statistically independent transition probabilities.  In preparation
for application to Bell-type experiments, we introduce a different
notation.  If system $A$ has input $\alpha$ and output $a$, and system
$B$ has input $\beta$ and output $b$, then the transition
probabilities are statistically independent, that is
$$
\Pr(a,b|\alpha,\beta) = \Pr(a|\alpha)\Pr(b|\beta).  \eqno(3.15)$$ Each
system has its transfer probabilities, $\Pr(F^A)$ and $\Pr(F^B)$.  These
are uncorrelated, so the corresponding transfer probability for the
combined system $S$ has the form
$$
\Pr(F)=\Pr(F^A,F^B) = \Pr(F^A)\Pr(F^B).
\eqno(3.16)$$

Using transfer functions we can distinguish between those correlations
that come entirely through the effect of correlated background
variables, and those that come through direct interaction between the
systems.  If there is no direct interaction, then there are no signals
between $A$ and $B$ and we can find transfer functions $F^A$ which are
functions of the input $\alpha$ alone and $F^B$ that are functions of
$\beta$ alone, as before,
$$
F^A(\alpha,\beta)=F^A(\alpha),\h10
F^B(\alpha,\beta)=F^B(\beta),
\eqno(3.17)$$
such that the probabilities for all other transfer functions are
zero.

However, if there is interaction through the background variables,
then these separate transfer functions are correlated, so that
$$
\Pr(F^A,F^B) \ne \Pr(F^A)\Pr(F^B)
\eqno(3.18)$$
In such a case there is correlation between the outputs, which is
produced by the action of the background variables on the systems $A$
and $B$, but it is not possible to use this correlation to send
signals between $A$ and $B$, because the probabilities of all
signalling transfer functions are zero.

There is also another possibility, which occurs for experiments of the
Bell type, for which there are signalling transfer functions
between ports at $A$ and $B$, which always have non-zero total
probability, yet it is not possible to use these to send signals,
since it is not known which transfer functions operate.  This is what
happens when quantum systems violate the Bell inequality.  Such
signals are called {\it weak signals}.  For a weak signal, the
stochastic processes in the system can only be explained in terms of
one or more non-zero signalling transfer probabilities, but it is not
possible for us to control the input in such a way as to send a signal
to the output corresponding to the signalling transfer function.  The
transfer functions are out of our control, as discussed using
different terms by Shimony \cite{Shimony1984}.

If the input and output of $A$ are both spatially separated from the
input and output of $B$, then according to classical special
relativity, there can be no signals from the input of $A$ (or $B$) to
the output of $B$ (or $A$), but there can be correlation between the
systems through interactions in their common past.  When these systems
are correlated, all the nonzero transfer probabilities satisfy (3.17)
and (3.18).  This is the locality condition, which can be applied to
quantum systems also, whether they are supposed to have background
(hidden) variables or not.

\vs{\bf 3.4 Linked systems and causal loops} 

Stochastic systems, like deterministic systems, can be combined by
linking inputs and outputs.  We do not go into the general theory of
such combinations, only the parts that are important for quantum
measurement theory.  

There are constraints on the combinations that are independent of
spacetime.  Corresponding to the conditions AD1 and AD2 for
deterministic systems, we have the stochastic system conditions

AS1.  A finite number $\N(S^k)$ of subsystems $S^k$, with fixed input
and output ports and joint transfer probability given by $\Pr(F^1,F^2,\dots
,F^{\N(S)})$, can be combined in any way to form a system $S$,
provided that the corresponding input and output ports are consistent.

AS2.  Combining systems does not change their joint transfer 
probability, given by $\Pr(F^1,F^2,\dots ,F^{\N(S)})$.

The most significant difference between these assumptions and those
for deterministic systems is that for the deterministic systems, we
could assume that the conditions can be applied to the individual
systems independently.  For stochastic systems, we can no longer do
this, because there can be correlations between transfer functions,
due to the classical or quantum environment, or to quantum
entanglement.  So we have to use joint probabilities.  For the
particular case of probabilities that are either 0 or 1, the
stochastic conditions become the same as the deterministic conditions,
so there are systems for which the stochastic conditions cannot be
satisfied, because they lead to forbidden causal loops.  

We now show that a stochastic causal loop can lead to contradictions
with finite probability under more general conditions.
 
Consider two elementary systems connected in series so that the output
port of $S^1$ is the input port of $S^2$.  The combined system has
only one input $i=i_1$ and one output $j=j_2$, which are illustrated
in figure 2.7.  A transfer function $F$ for the combined system is
obtained in terms of transfer functions $F^1$ and $F^2$ for the
individual systems as in equation (2.8) for deterministic systems as
$$
j = F^2(F^1(i)) = F(i),\hx{where}\h10 F=F^2\o F^1.
\eqno(3.19)$$
If there is no correlation through background variables, or
its quantum equivalent, then the probability $\Pr(F)$ of a given
transfer function $F$ for the combined system is
$$
\Pr(F) = \sum_{F^1,F^2}\Pr(F^1,F^2)\delta(F,F^2\o F^1).
\eqno(3.20)$$
When there is no correlation between the transfer functions, through
classical background variables or quantum entanglement, then
we can put 
$$
\Pr(F^1,F^2)=\Pr(F^1)\Pr(F^2),
\eqno(3.21)$$
a case of particular importance for the double Bell experiment
that is to follow.  In that case if, if $\Pr(F^1)\ne 0$ for $S^1$
and $\Pr(F^2)\ne 0$ for $S^2$, then $\Pr(F)\ne 0$ for the combined
system $S$.

Just as in the case of deterministic systems, a chain of stochastic
systems can be joined end to end to make a causal loop, with the
transfer functions given by equation (2.9).  For stochastic systems,
causal loops are allowed if there are loop transfer probabilities,
for which the probability of {\it every} forbidden loop transfer
functions is zero.  Otherwise there is a nonzero probability of a
contradiction.  So at least one of the assumptions must be wrong.

This is the {\it stochastic loop constraint}

An example of a forbidden stochastic causal loop can be constructed
from the deterministic example of figure 2.10, with transfer functions
given by equations (2.10).  All we have to assume is that the probability
$\Pr(F^k)$ for the transfer function $F^k$ of system $S^k$ is not zero,
and that the transfer functions are uncorrelated, so that
$$
\Pr(F^1,F^2,F^3,F^4) = \Pr(F^1)\Pr(F^2)\Pr(F^3)\Pr(F^4)
\eqno(3.22)$$ 
There is then a nonzero probability for this forbidden deterministic
loop transfer function, that is, a nonzero probability of a
contradiction.

So the corresponding stochastic causal loop is itself forbidden.

As we shall see, the stochastic loop constraint has an application in
quantum measurement.  This is only possible because the nonlocality of
quantum measurement imposes weaker spacetime constraints on causality
than for classical deterministic or stochastic systems, or, indeed,
for purely quantum systems.  

Spacetime constraints are introduced in the next section.

\vs\vs\cl{\bf 4 Space and time}

\vs{\bf 4.1 Time-ordering and causality}

So far we have considered the causal relations between inputs, systems
and outputs, without going into much detail on their spacetime
relations.  Does a particular input $i_p$ occur before or after
another input, or an output, or, in special relativity, are they
separated by a spacelike interval?  These spacetime relations between
the input and output ports are introduced quite separately from the
causal relations.  This is particularly important when applying the
theory to the subtle spacetime causal relations of quantum measurement
problems, like those of Bell experiments.  We do not assume {\it a
priori} that an output cannot be affected by an input that is
separated from it by a spacelike interval.  {\it The relations between
causality and spacetime must be introduced explicitly, and cannot be
assumed.}

In nonrelativistic Galilean dynamics, an input that takes place later
than an output cannot affect it.  This applies also in classical
(non-quantum) special relativity, if `later' than the output is
interpreted to mean `entirely in or on the forward light cone of every
spacetime point' of the output.  Each of these conditions forbids any
kind of causal loop.  Such relations between spacetime and causality
are discussed in this chapter.

Classical nonrelativistic mechanics assumes that there is a universal
time, represented by a variable $t$, defined everywhere in space.
Given a system S, with any number of input and output ports, with
input port $p_{\rm in}$ that comes after an output port $p_{\rm out}$,
then there are no signals from $p_{\rm in}$ to $p_{\rm out}$.  These
nonrelativistic relations are independent of the location of the ports
in space.

Classical relativistic dynamics is not so simple, since `before' and
`after' then apply only to events with timelike or null separation.
They are undefined for spacelike separations.  Classically there are
are no signals that go faster than the velocity of light, and so no
signals between input and output ports that have spacelike separation.
Relativistically, spatial relations become important.

Because the velocity of light is large, it is not always easy to
control the spacetime relations experimentally.  In particular, it is
not easy to make the intervals between two ports of a laboratory
experiment completely spacelike, because the time intervals are so
short.  This is one of the problems of experiments of the Bell type,
leading to the `locality loophole' which took a long time to close
\cite{Aspect1999,Weihs1998,Tittel1999}.

\vs{\bf 4.2 Systems and ports in space and time}

We consider a finite number $\N(S)$ of systems $S$.  Each system
occupies a finite region of spacetime, which we will also denote by
$S$, and a finite number of input ports and output ports.  The ports
of $S$ normally occupy regions of spacetime on the spacetime boundary
of $S$.  Systems can be connected through their input and output
ports.  Normally, they do not otherwise overlap in spacetime.  It is
sometimes convenient to show the spacetime relations of the ports in a
spacetime diagram.  Where the spacetime relations of special
relativity are important, lines at 45$^\o$ represent the velocity of
light.  These are seen in the figure 5.1 and figures 5.2, 5.3 and 6.1
of the Bell and double Bell experiments.

{\bf 4.3 Signalling transfer functions}

These can be considered independently of spacetime, but their
connection with the spacetime relations is so important that we treat
them here.

Sometimes the input and output $\alpha,a$ in spacetime region $A$ are
far from the input and output $\beta,b$ in spacetime region $B$, and
causal connections for deterministic systems can the be expressed in
terms of signals between the two regions, as follows:
$$
F^A(\alpha \beta)=F^A(\alpha),\hskip4mm F^B(\alpha \beta)=F^B(\beta)
\hskip6mm\hbox{(No signals between $A$ and B)}
\eqno(4.1a)$$$$
F^A(\alpha \beta)=F^A(\alpha),\hskip5mm F^B(\alpha \beta)\ne F^B(\beta)
\hx{(Signal from $A$ to $B$ only)}
\eqno(4.1b)$$$$
F^A(\alpha \beta)\ne F^A(\alpha),\hskip5mm F^B(\alpha \beta)=F^B(\beta)
\hx{(Signal from $B$ to $A$ only)}
\eqno(4.1c)$$$$
F^A(\alpha \beta)\ne F^A(\alpha),\hskip5mm F^B(\alpha \beta)\ne F^B(\beta)
\hskip3mm\hx{(Signals both ways),\hskip3mm}
\eqno(4.1d)$$
where, for example, the inequality in (4.1b) means that $b$ depends
explicitly on $\alpha$.  Those transfer functions that allow signals are
called {\it signalling} transfer functions.  The signals are
constrained by the conditions of special relativity, depending on
whether the intervals between the inputs and outputs are spacelike or
timelike.  Where there are no signals, there are no corresponding
signalling transfer functions, and no direct causal connection.
However than can be a causal connection between the systems through
classical background variables, or through interaction with the 
same quantum system.

A signalling transfer function for a signal from $A$ to $B$, where
$A+B$ is a deterministic system, has the property that there must be
at least two values of the input $\alpha$ of $A$, which we can call
$0$ and $1$, and two values of the output $b$ of $B$, which we can
call $+$ and $-$, such that when $\beta$ is constant, the input
$0$ gives the output $+$ and the input $1$ gives the output $-$.
 
Now suppose the systems are stochastic.  The signals and causal
connections are then expressed in terms of the transition or transfer
probabilities.  In that case, for all consistent transfer
probabilities, if there is a signalling transfer function for a signal
from $A$ to $B$ with nonzero probability, with arbitrary $\N(a)$ and
$N(b)$, the inputs can be restricted to a particular pair of values of
$\alpha$ and a particular value of $\beta$, such that for all
consistent transfer probabilities, the restricted transfer function
also has nonzero probability.  That is, the inputs can be restricted
to a binary input for $\alpha$ and a single, degenerate, input value
of $\beta$.  This is important for the simplified Bell experiment
discussed in section 5.5 and is used for the double Bell experiment of
section 6.2.

 \vs\vs\cl{\bf 5 Inequalities of the Bell type}

\vs{\bf 5.1 Quantum experiments in spacetime}

A quantum experiment consists of a quantum system linked to classical
systems through preparation events $\alpha,\beta,\dots$, which are the
inputs, and measurement events $a,b,\dots$, which are the
outputs. The input $i=(\alpha,\beta,\dots)$ (in the singular) represents
all the input events, and the output $j=(a,b,\dots)$ all the
output events.  Classical systems can similarly be prepared by input
events and measured by output events.

Special relativity puts strong constraints on causality in spacetime.
According to classical special relativity, causal influences such as
signals operate at a velocity less than or equal to the velocity of
light, so that an input event influences only output events in or on
its forward light cone, and an output is only influenced by events in
or on its backward light cone.  They cannot operate over spacelike
intervals, nor can they operate backwards in time.  There is only {\it
forward causality}.

Nevertheless, there can be correlations between systems with spacelike
separation, due to background variables that originate in the region
of spacetime common to their backward light cones.  Causality
which does act over spacelike intervals is generally called nonlocal
causality, or simply {\it nonlocality}. It does not exist for
relativistic classical systems.  Correlation between events with
spacelike separation is not sufficient evidence for nonlocality, since
it can be due to common background variables.

Causality which acts backwards over timelike intervals means that the
future influences the present, and the present influences the past.
This is {\it backward causality}.  This term will only be used when
there is timelike separation between cause and effect.  For our
systems, with classical inputs and outputs, there can clearly be no
backward causality, because it would be easy to connect a system with
backward causality to an ordinary classical system with forward
causality to produce a forbidden causal loop, using the method
described in detail for the double Bell experiment of section 6.2.
However, for purely quantum systems, we have no clear evidence either
way as to whether or not there can be backward causality, and this
presents a problem for Hardy's original derivation of his theorem.

For our systems, these relations can be expressed in terms of
classical inputs and outputs.  Suppose that the input events or inputs
$\alpha,\beta,\dots$ and output events or outputs $a, b,\dots$ are
all so confined in spacetime that the interval between any pair of
such events is well-defined as spacelike or timelike.  The special
case of null intervals is excluded, as it is not needed for our
purposes.  Since no signal can propagate faster than the velocity of
light, an output $j_\Y$ depends only on those $i_\X$ that lie in its
backward light cone, so that any variation of the other $i_\X$ makes
no difference to $j_\Y$.  This is illustrated in figure 5.1.  This
condition on the transfer function $F$ is a condition of forward
causality, which applies to all those deterministic classical and
quantum systems for which the input uniquely determines the output.
\begin{figure}[htb]

\epsfxsize14.0cm
\centerline{\epsfbox{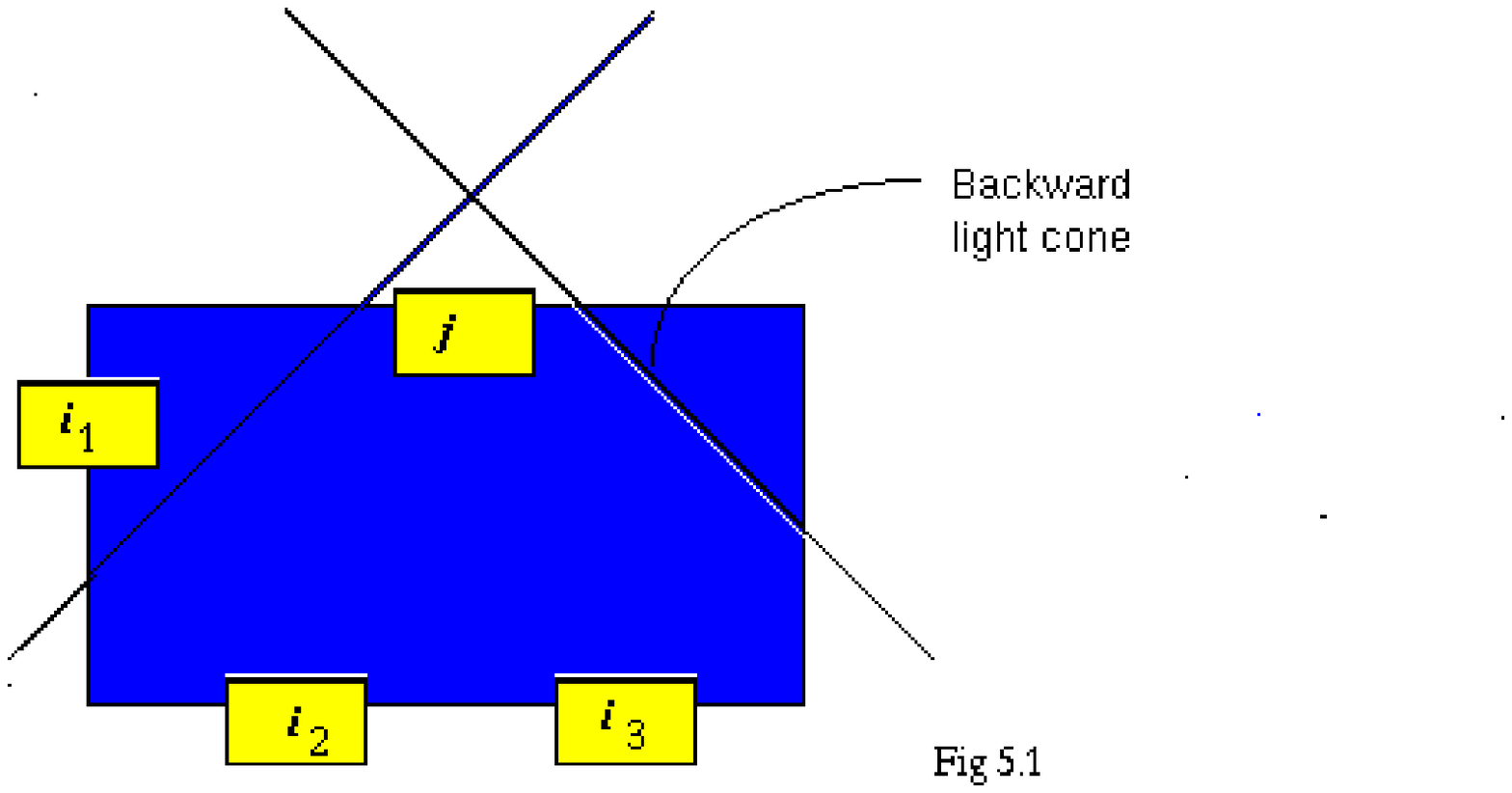} }
\label{fig5.1}

\caption{Classical system $S$ with spacetime relations of inputs and 
outputs.  The output $j$ can depend on $i_2$ and $i_3$, but not on $i_1$}
\end{figure}

\vs{\bf 5.2 Background variables}

The background variable $\lambda$ of noisy classical systems may be
considered as an additional input, so it does not affect the causality
relations between the inputs and outputs.  Each of the transfer
functions $F_\lambda$ satisfies the same causality conditions as the
transfer function $F$ of a deterministic system.  There is only
forward causality.

The same applies to the transfer functions $F_\lambda$ of some quantum
systems, but not to all.  According to quantum mechanics, there are
experiments for which the $F_\lambda$ with hidden variables $\lambda$
do not satisfy the conditions of forward causality: causality is
nonlocal.  These include experiments designed to test the violation of
Bell's inequalities.  Experimental evidence has been overwhelmingly in
favour of quantum mechanics, and also favours nonlocal
causality through violation of inequalities.  Nonlocal causality implies
weak nonlocal signals in the sense of section 3.3. 

The example of Bell experiments shows that the presence or absence of
weak signals can tell us something about the properties of quantum
systems.  Because the background variables are hidden, weak signals
cannot be used to send signals faster than the velocity of light.
However, systems with weak signals have observable properties that
cannot be simulated by any system whose inputs and outputs are linked
by classical systems only.

\vs{\bf 5.3 Bell-type experiments}

Einstein-Podolsky-Rosen-Bohm experiments, and Bell experiments which
test Bell or Clauser-Holt-Horne-Shimony inequalities, are typical
quantum nonlocality thought experiments.  Real experiments have been
based on the CHHS inequality, and use polarized photons.  The
analysis is similar for the Bell inequality and spin-half particles in
magnetic fields, which we treat here.  In one run of an experiment,
two such particles with total spin zero are ejected in opposite
directions from a central source.  This stage of the preparation is
not included in the input, as it happens for every run of the
experiment, and the inputs $i$ are normally variable.  Before the
particles are detected, they pass into Stern-Gerlach magnetic fields.
The classical settings of the orientations of the magnetic fields are
the input events.  The timing of these settings is crucial.  The
measurements, which include the classical recording of one of the two
spin directions parallel or antiparallel to the field, are the output
events.
\begin{figure}[htb]

\epsfxsize14.0cm
\centerline{\epsfbox{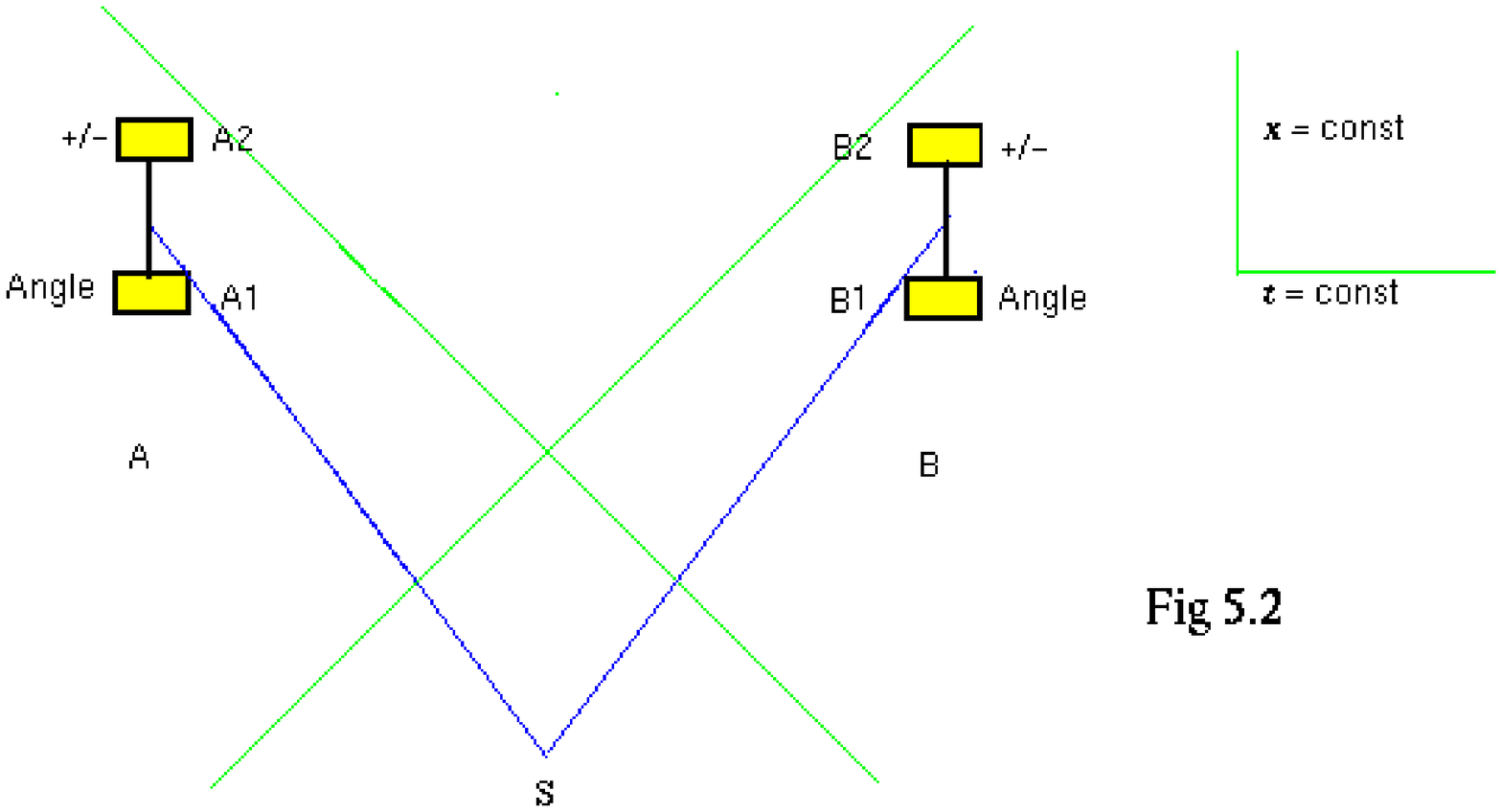} }
\label{fig5.2}

\caption{Spacetime diagram of Bell's experiment. Thin diagonal lines
at $45^o$ represent the velocity of light.  At both $A1$ and $B1$, the
angle is the setting of the angle of spin or polarization measurement,
which is an input event, and at $A2$ and $B2$, $+/-$ represents the recording of
the spin or polarization, an output event.}
\end{figure}

A spacetime illustration of a Bell experiment to test Bell's
inequality is given in figure 5.2.  There are two inputs and two
outputs, given by
$$
i=(\alpha, \beta),\h10 j=(a, b), 
\eqno(5.1)$$ 
where one pair of inputs and outputs is confined to a spacetime
region $A$ and the other is confined to a spacetime region $B$.  The
output $a$ is in the forward light cone of $\alpha$ and similarly for
$b$ and $\beta$.  The input preparation at $A$ must be separated by a
spacelike interval from the output measurement at $B$ and vice versa,
so both the input and output at $A$ are spatially separated from both
the input and output at $B$.  This is a considerable experimental
challenge.

For local hidden variables, it is not possible to send even weak
signals between $A$ and $B$, so there are only transfer functions of
type (4.1a), and the other 3 types of transfer function have
probability zero.  It follows that the only transfer functions with
nonzero probability have the form
$$
F = (F_A,F_B), \hx{where} \hskip5mm
a = F_A(\alpha),\hskip5mm b = F_B(\beta).
\eqno(5.2)$$
The conditional probability for the outcome $a,b$ given
inputs $\alpha,\beta$ is then
$$
\Pr(a,b|\alpha,\beta) = \sum_{F_A,F_B}
\Pr(F_A,F_B)\delta(a,F_A(\alpha))\delta(b,F_B(\beta)).
\eqno(5.3)$$
This has the same form as Bell's condition for locality, but
in terms of transfer functions instead of hidden variables.

The spin of a particle is denoted $+$ when it is in the same direction
as the field and $-$ when it is in the opposite direction.  The output
$j=(a, b)$ then consists of one of four pairs of spin components,
$(++)$, $(+-)$, $(-+)$ and $(--)$, where the first sign represents the
output at $A$ and the second sign the output at $B$.  We will adopt a
convention whereby the angles at $A$ and $B$ are measured from zeros that
point in opposite directions, so that, for total spin zero, when the
angles are equal, the measured signs of the spin components are the
same, not opposite.

Consider first an experiment for which each of the magnetic field
orientations at $A$ and at $B$ has only two possible angles,
$\theta_i$ with $i=1,2$, the same pair of angles at $A$ and $B$ , with
the above convention.  Then the function $F_A$ can be labelled by a
table of its output values $(\pm)$ for $i=1$ and $i=2$, and similarly
for $F_B$.  Thus if $a=+$ for $\alpha = 1$ and $a=-$ for
$\alpha=2$, we can denote $F_A$ by $F_A=[+-]$.  The function $F$ can
then be labelled by a table of 4 values, those for $A$ first, and those
for $B$ second, eg $[+-,+-]$.  For general functions $F$ we would have
$2^4=16$ output values.  But since we are assuming locality, $F$ has
the form (5.3), and only 2 arguments each are needed to specify $F_A$
and $F_B$, so we need only 4 output values to specify $F$.

The number of independent probabilities $\Pr(F)=\Pr(F_A,F_B)$ is
then severely restricted by the fact that for opposite input
orientations of the magnetic fields at $A$ and $B$, corresponding to
$\theta_A=\theta_B$ and $\alpha=\beta$, the output signs $a,b$
must be the same.  So when they are different the probability of $F$
is zero.  Reversing the direction of all spins does not affect the
probabilities.  Thus all $\Pr(F)$ are zero except
$$
P_1 = \Pr([++,++])  = \Pr([--,--]),
\eqno(5.4a)$$$$
P_2 = \Pr([+-,+-])  = \Pr([-+,-+]),
\eqno(5.4b)$$
where the right-hand equalities follow by symmetry.  Notice that the
two strings of signs, for $A$ and for $B$, are the same.
\page

\vs{\bf 5.4 Bell inequalities}

There are no contradictions with only two directions for the magnetic
field.  They can be handled with local hidden variables.  But now
consider 3 directions $\theta_1,\theta_2,\theta_3$, as in a Bell
inequality.  The values of $\alpha$ and $\beta$ are 1,2,3, and there are
four independent probabilities, given by
$$
P_0 = \Pr([+++,+++]) = \Pr([---,---]),
$$$$
P_1 = \Pr([-++,-++]) = \Pr([+--,+--]),
$$$$
P_2 = \Pr([+-+,+-+]) = \Pr([-+-,-+-]),
$$$$ 
P_3 = \Pr([++-,++-]) = \Pr([--+,--+]).
\eqno(5.5)$$
From these transfer probabilities, we can obtain the known transition
probabilities for experiments in which $\alpha,\beta$ with $i=1,2,3$
correspond to the three orientations $\theta_i$ of the fields.
$\Pr(a b|\alpha \beta)$ is given by adding together the
probabilities for which $a$ appears in position $\alpha$ in the first
sequence of signs, and $b$ appears in position $\beta$ in the second
sequence:
$$
\Pr(++|11) = \Pr(--|11) =  P_0 + P_1 + P_2 + P_3= 1/2 \hx{and cyclic,}
\eqno(5.6a)$$$$
\Pr(++|23) = \Pr(--|23) = P_0 + P_1  \hx{and cyclic,}
\eqno(5.6b)$$$$
\Pr(+-|23) = \Pr(-+|23) = P_2 + P_3  \hx{and cyclic,}
\eqno(5.6c)$$
where the cyclic permutations permute 1,2 and 3 but not 0.

The equations can be solved for the $P_k$, all of which must be
nonnegative.  Their solutions, with resultant conditions on
$\Pr(++|\alpha \beta)$, $\Pr(+-|\alpha \beta)$ are
$$
2P_0 = \Pr(++|23)+ \Pr(++|31)+ \Pr(++|12) -{1\over 2} \ge 0 \hx{and
cyclic,} \eqno(5.7a)$$$$ 2P_1 = {1\over 2}+ \Pr(++|23)- \Pr(++|31)-
\Pr(++|12) \ge 0 \hx{and cyclic,} \eqno(5.7b)$$$$ 2P_1 = \Pr(+-|31)+
\Pr(+-|12)- \Pr(+-|23)\ge 0 \hx{and cyclic}.  \eqno(5.7c)$$ The
inequalities (5.7a,b) are always satisfied, but (5.7c) gives the Bell
inequalities \cite{Bell1987a}, chapter 2, his equation (15).  Bell
uses the same origin for the angles at $A$ and $B$, so his opposite
signs are the same signs here, and his expectation values of products
are
$$
P({\bf b},{\bf c}) = 4\Pr(+-|23) - 1  \hx{and cyclic}.
\eqno(5.8)$$
The condition of weak nonlocality, the condition that the transfer
function probabilities should all be non-negative, and the condition
that sums over them are all less than or equal to 1, provide a
systematic method for obtaining inequalities of the Bell type.

Inequalities of the Bell type follow from the condition that all
transfer functions have positive or zero probabilities, which suggests
the use of the methods of linear programming \cite{Percival1998b}.

Since quantum mechanics violates the Bell inequalities, it is weakly
nonlocal, and there is at least one weak signalling or nonlocal
transfer functions $\Fs$ or $F_{\rm NL}$ with nonzero probability
$\Pr(\Fs)$.  Using the notation of figure 5.2, there is a weak signal
from $A1$ to $B2$ or from $B1$ to $A2$, or both.

It is here that special relativity comes in.  The experiment is
invariant for a rotation of $\pi$ that interchanges $A$ and $B$.  This
rotation is an element of the Lorentz group.  So if there is a non-zero
probability of a weak signal in one direction, it follows from
invariance under this element of the group that there must also
be a non-zero probability for a weak signal in the other.  By Lorentz
invariance there must be weak signals in both directions.  The
violation of the Bell inequality is not enough on its own to ensure
this.

We will use this Lorentz invariance as an additional (spacetime)
assumption for stochastic systems:

AS3 For flat spacetime, the laws of physics are invariant under
all elements of the Lorentz group.

\vs{\bf 5.5 Simplified and moving Bell experiments}

The proof of Bell's theorem required at least three orientations of
the polarizers, but its consequences can be expressed in terms of the
transfer function of a {\it simplified Bell} experiment, using the
results presented at the end of section 4.3.  The spacetime
configuration is the same as previous two experiments, illustrated in
figure 5.2.  But the input at $A$ has only one value $\theta_1$.  The
input at $B$ has two values:
$$
\theta_1:\hs i=1,\h10 \theta_2:\hs i=2.
\eqno(5.9)$$
There are four possible output values $j$ given by
$$
(a+,b+)=\hs ++,\hskip3mm (a+,b-)=\hs +-,\hskip3mm (a-,b+)=\hs
-+,\hskip3mm (a-,b-)=\hs --, 
\eqno(5.10)$$ 
where the first sign is for $i=1$ and the second for $i=2$, and where
the second short form with just the two signs is used hereafter.

If $i=1$, corresponding to the same angle at $A$ and $B$, measured, as
usual, from zeros that are an angle $\pi$ apart, then $j=++$ or $j=--$.
Consider the $++$ case.  There are four possible transfer functions.
For the local degenerate transfer functions, the sign
at $A$ is independent of the angle at $B$, so $F_1$ and $F_4$ are local.  
For nonlocal (NL) or signalling transfer functions, the sign at A
depends on the angle at $B$, so $F_2$ and $F_3$ are nonlocal.

By Bell's theorem and Lorentz invariance, both of the nonlocal
transfer functions have non-zero probability:
$$
\Pr(F_2) >0,\h10  \Pr(F_3) > 0.
\eqno(5.11)$$

This simplified Bell experiment, like the full Bell experiment, is
depicted in figure 5.2 at rest with respect to a laboratory frame.
Now we make the whole experiment move uniformly, so that it is at rest
in a frame that is moving with respect to the laboratory frame, as
represented in figure 5.3.  In this figure $x$ and $t$ are position
and time coordinates in the moving frame.

\begin{figure}[htb]

\epsfxsize14.0cm
\centerline{\epsfbox{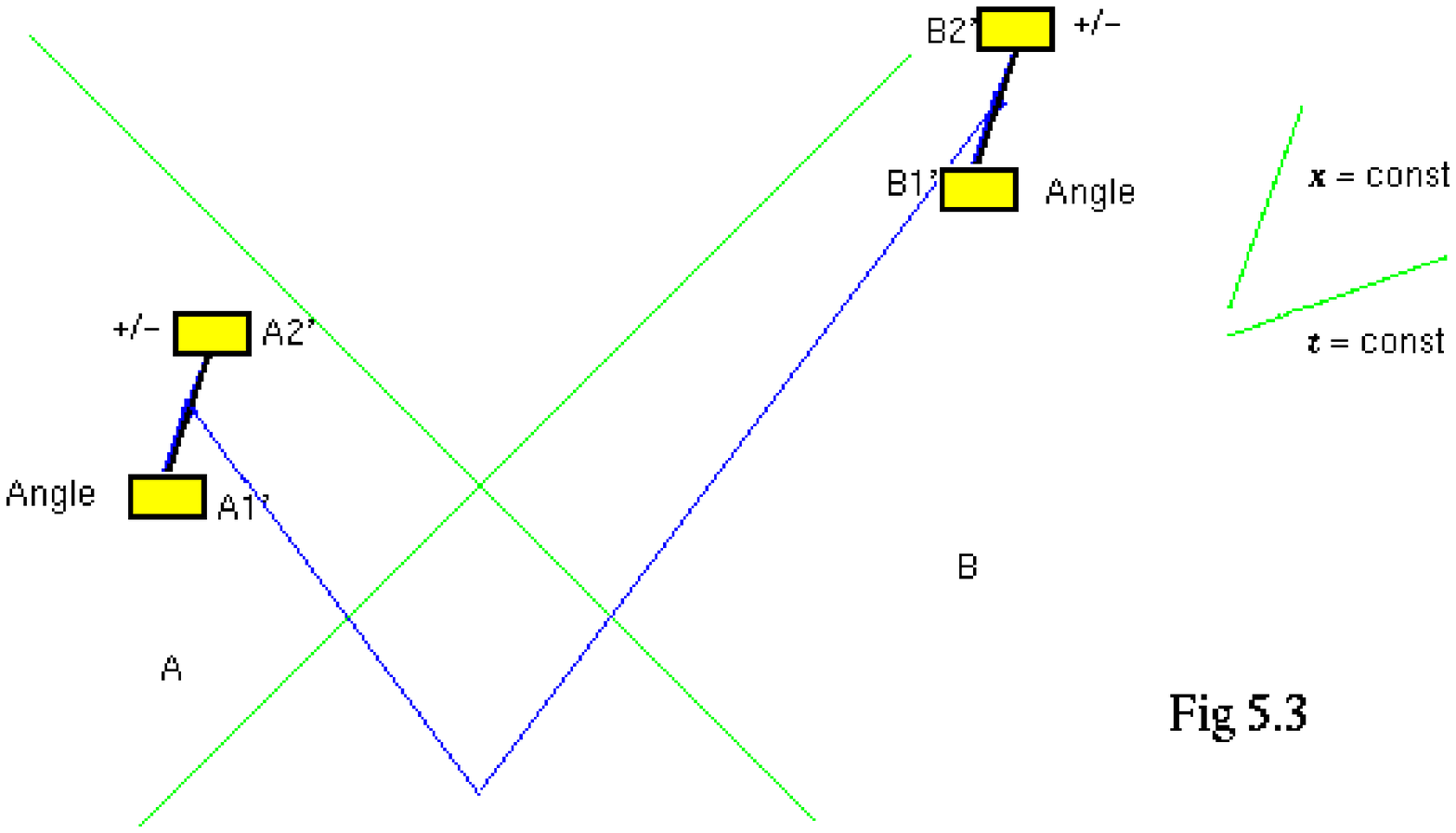} }
\label{fig5.3}

\caption{Spacetime picture of moving Bell experiment}
\end{figure}

The simplified moving Bell experiment is one component of the double
Bell experiment which follows.  Another component is a second
simplified moving Bell experiment which moves with equal speed in the
opposite direction, whose spacetime picture is a reflection of figure
5.3 in a vertical axis.

\vs\vs\cl{\bf 6 Hardy's theorem and the double Bell experiment}

\vs{\bf 6.1 Hardy's theorem}

Hardy's theorem \cite{Hardy1992a} goes further than Bell's theorem on
nonlocality of quantum measurement.  It states
that any dynamical theory of measurement, in which the results of the
measurements agree with those of ordinary quantum theory, must have a
preferred Lorentz frame.  The theorem does not determine this frame.

Hardy originally presented the theorem through a thought experiment
involving two matter interferometers, one for electrons and one for
positrons, with an intersection between them that allows annihilation
of the particles to produce gamma rays.  The proof of the theorem
depends on an assumption, which is that there is no backward causality
in quantum systems.  As discussed in section 5.1, we have no hard
evidence for or against such backward causality.  On the other hand,
given a choice between backward causality in quantum systems and the
breakdown of Lorentz invariance, it is not so clear which distasteful
choice we should make.  Without the assumption, this proof breaks
down.

A different derivation based on classical links between two Bell
experiments is given in \cite{Percival1998b,Percival1998d}. This
is the double Bell experiment.  The
proof of Hardy's theorem using the double Bell experiment is
independent of any assumptions about causality in the quantum domain.
It depends on causal relations between classical inputs and outputs,
using the transfer picture, and the stochastic loop constraint of
section 3.  It is also assumed that the results of the separate Bell
experiments are not correlated through some unsuspected background
correlation.  It would be remarkable if it were present, but a
more complicated {\it multiple} Bell experiment shows that the 
theorem is independent even of this assumption.  The multiple Bell
experiment is described in \cite{Percival1998b}, but not here.

\vs{\bf 6.2 The double Bell experiment}

Einstein's old simultaneity thought experiment used classical light
signals from a source and two pairs of receivers, each pair in a
different moving frame.  We replace this by two simplified moving Bell
experiments, each in a different frame and then show that the
assumption of invariance under Lorentz transformations, together with
further assumptions that are discussed in detail, leads to a forbidden
causal loop with nonzero probability.  So, given the further
assumptions, there is no Lorentz invariance.

Here is the argument.  It does not assume weak locality. 

The double Bell experiment is a combination of two simplified moving Bell
experiments, as defined at the end of the previous section, with two
classical links, as illustrated in figure 6.1
\begin{figure}[htb]

\epsfxsize14.0cm
\centerline{\epsfbox{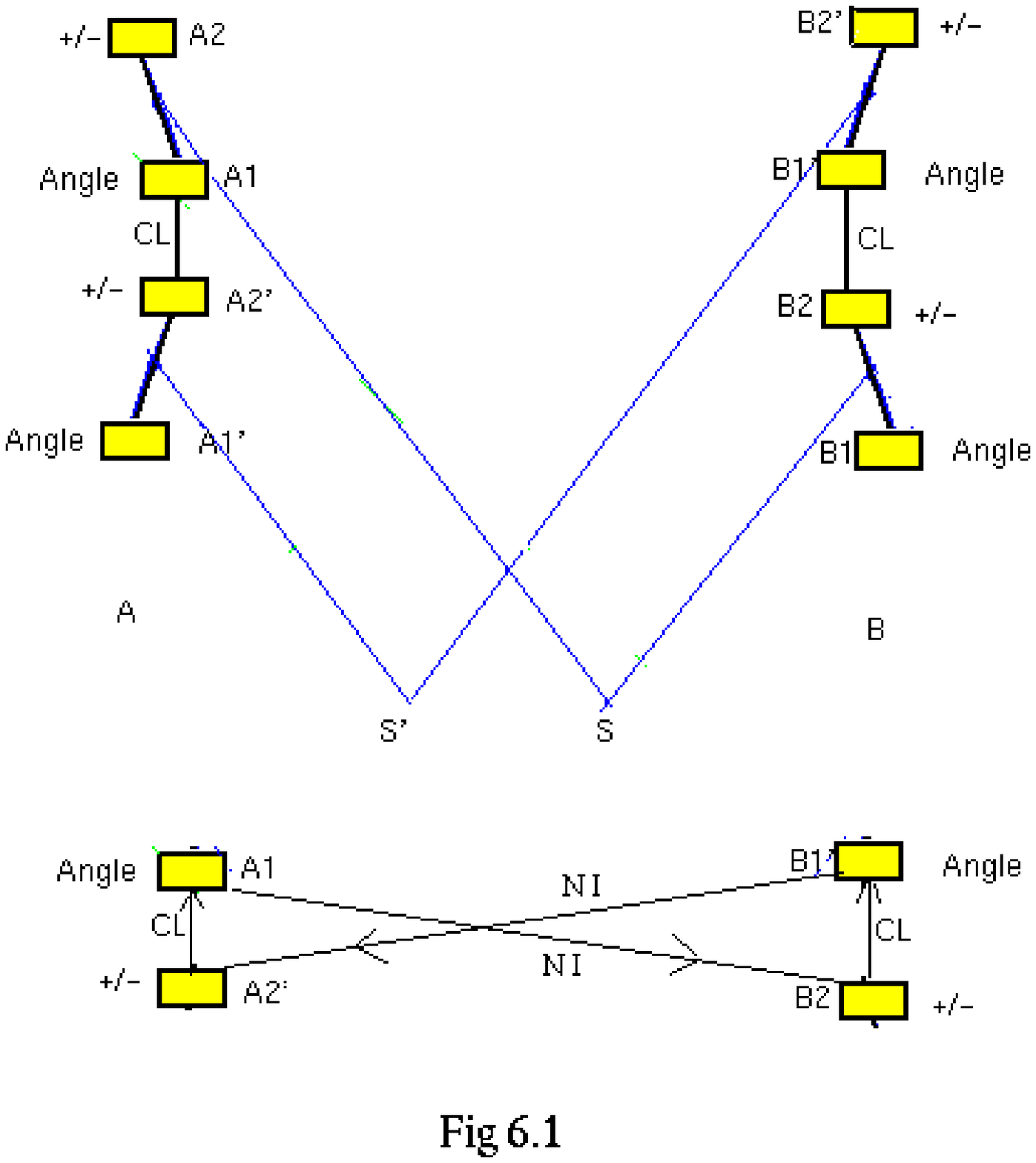} }
\label{fig6.1}

\caption{Spacetime diagram of the double Bell experiment.  The meaning
of the symbols corresponds to their meaning in the figures 5.2 and
5.3.  The primed symbols belong to one Bell experiment, and the
unprimed to the other.  As usual $A$ and $B$ represent spatially
separated parts of the full experiment, so there are four ports on
each side.  Notice that both the ports of {\it each} experiment at
$A$ are spatially separated from both the ports of each experiment at
$B$.  CL are classical links, by which an output from each experiment
determines the input to the other.}
\end{figure}

The two Bell experiments are independent at the quantum level, but
they are linked classically so that the output $A2'$ controls the
input $A1$, which is in its future lightcone, and the output $B2$
controls the input $B1'$.  There are two significant nonlocal
interactions NI.  One is the setting of the angle at $A1$ affecting
the measurement at $B2$, and the other is the setting of the angle at
$B1'$ affecting the measurement at $A2'$.  This is an unusual thought
experiment, in that the variable input orientations of the
magnetic fields at $A1$ and B1$'$ are determined entirely by outputs.
The causal loop structure is clearly seen in the laboratory.

The causal loop is 
$$
{A2}'\CL{A1}\NI{B2}\CL{B1}'\NI{A2}',
$$
where NI represents the nonlocal interaction of one or other of the
Bell experiments, and CL represents one of the classical links.
Section 5.5 on the simplified moving Bell experiment, shows that both
the nonlocal transfer functions must have nonzero probability.  All
inputs and outputs are binary, and with an appropriate convention for
the labelling inputs and outputs, the nonlocal interactions are both
equivalent to the identity gate $\I$.  We are free to choose the
classical links at will.  We choose the link at $A$ to be the
identity, and the link at $B$ to be the $NOT$ gate.  The loop transfer
function is then a $NOT$ gate, so it is a forbidden causal loop, and
its probability is not zero.

This is forbidden by the stochastic loop constraint, so AS1, AS2 or
the Lorentz assumption AS3 of section 3.4 must be false.  Since there
is nothing to prevent us from combining the two Bell experiments and
the two classical links together in the configuration of a double Bell
experiment, AS1 is satisfied, so the problem must be AS2 or Lorentz,
and we had better look very carefully at AS2: combining systems does
not change their joint transfer probability.

Two of the systems are deterministic gates, the other two are
stochastic single Bell experiments.  Before they are combined
they have independent transfer probabilities, so according to AS2,
this independence should be preserved when they are combined to
make the double Bell experiment.

It is a fundamental assumption of modern experimental physics that we
can control and monitor physical systems using deterministic digital
gates, with well-defined ports, such that the only interaction between
another system and a gate is through its ports.  If this were not true
we could not reliably use digital circuits for monitoring and
controlling physical systems.  So unless we are prepared to question
the basis of almost all modern experimental physics, the problem
cannot be AS2 as it applies to the interaction between the classical
gates and the single Bell experiments, and between one gate and
another.

The only remaining possibility is that AS2 should be wrong because the
joint transfer probabilities of the two Bell experiments are changed
as a result of combining them through the gates.  This seems highly
unlikely and appears to be ruled out by the multiple Bell experiment
of \cite{Percival1998b}.

Otherwise we have to conclude that AS3 is wrong,  that in flat
spacetime, the laws of physics are not invariant under the operations
of the Lorentz group.

The double Bell experiment is described here as a thought experiment.
It might be worth performing as a real experiment, but this is not
clear.  The results of the double Bell thought experiment follow from
the results of an ordinary single Bell experiment, and some very
general assumptions about independence of systems.  These assumptions
are at the foundation of almost all modern experimentation.  The
assumptions can be checked directly, and more simply, without
performing the double Bell experiment itself.

However it becomes even more important than before to close the final
detection loophole in the single Bell experiment, since it is assumed
in our derivation that inequalities of the Bell type are violated with
the right spacetime relations between the input and output ports, and
this has not yet been fully confirmed experimentally, because of the 
loophole.  

The double Bell experiment as presented here would be difficult to
perform in the laboratory, because the apparatus of the two single
Bell components would have to be move fast enough to produce a time
dilation greater than the summed delays between the inputs and outputs
of the Bell experiments and classical links.  A more practical
static double Bell experiment is presented in
\cite{Percival1998b,Percival1998d}, using optical fibre delay lines.
Unfortunately, it was not realized when that paper is written, that
the introduction of the time delays destroys the $\pi$-rotation
symmetry of its component single Bell experiment, so it can't be used
to invalidate Lorentz invariance without additional assumptions. 

It might be interesting to do the static experiment, even if it does
produce the results expected from ordinary quantum theory, because it
might have a weak causal loop, and no experiment to date has been set
up with even the possibility of such a loop.  Like the moving
experiment the static experiment has no classical input in the usual
sense, because the inputs of the individual simplified Bell
experiments are controlled entirely by two of their outputs.  Hence
the loop.

\vs\vs\cl{\bf 6.3 Breaking Lorentz symmetry}

We conclude that quantum measurement in general, and Bell nonlocality
in particular, break Lorentz symmetry.

Claimed Lorentz-invariant realistic theories of quantum measurement
\cite{Breuer1998,Berndl1996,Samols1995,Ghirardi1990a} and also
\cite{Bell1987a}, chapter 22, appear to conflict with Hardy's theorem
and our conclusions.  It would be interesting to apply these theories
to the double Bell experiment to find out how.  For some of these
theories the apparent conflict may lie in the different meanings given
to `Lorentz invariant' as it applies to quantum measurement theory.

Any localized system, any macroscopic system, any planet, breaks
Lorentz symmetry, and causes other systems in its neighbourhood to
obey a dynamics with a preferred frame.  This is not unusual, it is
the general rule for both classical and quantum localized systems.  We
can restore Lorentz symmetry for a system $S$ by enlarging $S$ to
include the system that breaks the symmetry.  Conversely, if we find a
dynamical behaviour that breaks the symmetry, the first thing to do is
look for a localized system in the environment that causes it.  This
could be provided by a field, as discussed for quantum measurement in
\cite{Percival1998d}.

But {\it we} need to break Lorentz symmetry in such a way that weak
signals can propagate faster than the velocity of light, which is not
true for the symmetry breaking due to ordinary localized systems.
This situation is similar to the symmetry-breaking of quantum field
theory in at least one respect:  processes that are forbidden when the
symmetry is respected become possible when it is broken.  These
issues, and the necessity of a consistent definition of simultaneity,
are discussed in \cite{Percival1998d}, but there is not yet an
adequate theory of such Lorentz symmetry-breaking.

\page

\vs{\bf Acknowledgements} I thank the Group of Applied Physics of the
University of Geneva and the Institute of Physics and Astronomy of the
University of Aarhus for their hospitality, also Gernot Alber, Lajos
Di\'osi, George Ellis, Nicolas Gisin, Lucien Hardy, Klaus M\/olmer,
Sandu Popescu and a large fraction of the Physics Department at QMW
for stimulating discussions.  I also thank the UK EPSRC and the
Leverhulme Trust for financial support during the period of the
research.

\vs{\bf Notation}

$S$ system, $i$ input value, $j$ output value, $\N(n)$ number of possible values
of $n$, $p$,$q$ port labels, $i_p,j_q$ input or output value at input or
output port.  $i\to j$ transition.   $F$ transfer function. $k$ system
or transfer function label.  S$^k$ $k$th system.  $F^k$ transfer function
for system S$^k$.  $F_k$ one of several transfer functions for a single
system,  $F_{\rm loop}$ loop transfer function.

An alternative notation for systems is $A$, $B$, and corresponding
inputs $\alpha$, $\beta$ and outputs $a$, $b$.

\vskip20mm
\bibliography{../bib-/icp}   
\bibliographystyle{abbrv}   

\end{document}